\documentclass[twocolumn,prd,showpacs,preprintnumbers,amsmath,amssymb,floatfix]{revtex4}

\usepackage{graphicx}
\usepackage{dcolumn}
\usepackage{bm}

\newcommand{\Slash}[1]{\ooalign{\hfil/\hfil\crcr$#1$}}
\newcommand{\re}{\text{Re}}

\newcommand{\ket}[1]{| \, #1 \, \rangle}

\newcommand{\piK}{\pi^- p\to K^-\Theta^+}
\newcommand{\Kpi}{K^+ p\to \pi^+\Theta^+}
\newcommand{\tenbar}{\overline{\bm{10}}}

\begin{document}

\preprint{}

\title{Phenomenological study of two-meson couplings of $\Theta^+$}

\author{Tetsuo~Hyodo}
\email{hyodo@rcnp.osaka-u.ac.jp}
\affiliation{%
Research Center for Nuclear Physics (RCNP),
Ibaraki, Osaka 567-0047, Japan
}%

\author{Atsushi~Hosaka}%
\affiliation{%
Research Center for Nuclear Physics (RCNP),
Ibaraki, Osaka 567-0047, Japan
}%

\date{\today}
\begin{abstract}
    We evaluate two-meson couplings of $\Theta^+$,
    using experimental information of nucleon resonances 
    decaying into $\pi\pi N$ channels, in which the two 
    pions are in scalar- and vector-type correlations.
    We examine two assignments of spin and parity of $J^P=1/2^+$ 
    and $3/2^-$, for which the experimental spectra of known 
    resonances with exotic baryons are properly reproduced by an
    octet-antidecuplet representation mixing scheme. With the 
    obtained coupling constants, total cross sections of the 
    reactions $\pi^- p \to K^- \Theta^+$ and $K^+ p\to \pi^+ \Theta^+$
    are calculated. Substantial interference of two terms may occur
    in the reaction processes for the $J^P=1/2^+$ case, whereas the 
    interference effect is rather small for the $3/2^-$ case.
\end{abstract}
    
%
\maketitle

\section{Introduction}\label{sec:intro}

Evidence for the existence of the exotic baryon 
$\Theta^+$~\cite{Nakano:2003qx} has been stimulating theoretical 
and experimental studies in hadron spectroscopy. The study of 
multiquark states in QCD will eventually lead to the understanding
of the rich structure of hadronic matter. Recently, however, the
existence of the $\Theta^+$ has become controversial~\cite{Hicks:2005gp}.
There are more than 10 experiments that indicate evidence of the
$\Theta^+$, whereas a similar number of experiments claim null results.
In such a situation, confirmation of existence (or nonexistence) of 
the $\Theta^+$ is urgent and crucially important. For this, it is 
strongly desired to clarify the reaction mechanism for the production
of $\Theta^+$. Among various possibilities, a particularly interesting
property that is expected to be characteristic to exotic baryons is 
their strong coupling to two-meson states in transitions to an ordinary 
baryon. This is the subject we study in the present paper.  

Studying two-meson coupling is important for several reasons. First, 
a heptaquark model has been proposed in the early stage of development 
to explain a light mass and a narrow decay width~\cite{Bicudo:2003rw,
Kishimoto:2003xy,Llanes-Estrada:2003us,Bicudo:2004cm,Huang:2004ti}.  
Although a quantitative study---in particular with a model of hadrons 
where $\Theta^+$ is regarded as a bound state of $\pi K N$ system---does
not work with the present knowledge of hadron interactions, a two-meson 
contribution to the self-energy of $\Theta^+$ has been shown to be
consistent with the expected pattern of the masses of the antidecuplet
members~\cite{Hosaka:2004mv}.

Second, the importance of two-meson coupling has been implied from an
empirical observation of the extended OZI rule~\cite{Roy:2003hk}. The
dominance of connected quark lines favors creation of a $q \bar q$ pair
in the transition of $\Theta^+(qqqq\bar q) \to N(qqq)$, which is
naturally associated with the coupling to two mesons, whereas a coupling
to a single meson is suppressed.  

Finally, two-meson coupling plays an important role in reaction studies.
Without two-meson coupling, all the amplitudes for $\Theta^+$ production
are proportional to the $\Theta^+KN$ coupling, which is fixed by the 
very small decay width of the $\Theta^+$. However, two-meson coupling
is determined from other source as we will see in the following, 
independently of the $\Theta^+KN$ coupling. Therefore, even with the
extremely narrow width of $\Theta^+$, a sizable cross section can be
obtained with two-meson coupling.
 
In Ref.~\cite{Hosaka:2004mv}, an analysis of the two-meson coupling is
performed in the study of the self-energy of the $\Theta^+$, assuming
that $J^P=1/2^+$ with $N(1710)$ in the same antidecuplet
($\bm{\overline{10}}$). Since the $\Theta^+$ cannot decay into 
$K\pi N$ channel, the coupling constants are determined from the $N^*$
decay into the $\pi\pi N$ channel and flavor SU(3) symmetry. Two types
of Lagrangians are found to be important for the self-energy of the 
baryon antidecuplet. It is also shown that the two-meson contribution 
is indeed dominant over a single-meson contribution. However, the 
assumption of pure $\bm{\overline{10}}$ may not be the case in reality.

This point is clarified in Ref.~\cite{Hyodo:2005wa}, where we study 
the phenomenology of flavor partners for the $\Theta^+$. We assign 
the masses of experimentally known particles in an octet-antidecuplet
mixing scheme, finding good fits for $J^P=1/2^+$ and $3/2^-$. The
decay width of the $\Theta^+$ is also evaluated in the same scheme,
and the $J^P=3/2^-$ case naturally explains the narrow width, in 
accordance with the quark model estimation~\cite{Hosaka:2004bn}. In 
both $J^P$ cases, we obtain relatively large mixing angles, which 
implies the importance of the representation mixing.

Hence, combining these two findings, we calculate the two-meson 
couplings including the representation mixing. First we determine the
coupling constants of $N^*\to \pi\pi N$ from the experimental widths 
and separate the $\bm{\overline{10}}$ component from the $\bm{8}$ 
component. Then, by using SU(3) symmetry, the coupling constants of 
$\Theta  K\pi N$ can be determined for $J^P=1/2^+$ and $3/2^-$,
including representation mixing of $\bm{8}$ and $\bm{\overline{10}}$.
We focus on the decay channels in which the two pions are correlated
in scalar-isoscalar and vector-isovector channels, which are the main 
decay modes of the resonances and play a dominant role in the $\Theta^+$
self-energy~\cite{Hosaka:2004mv}.

As an application of the effective Lagrangians, we perform the 
analysis of the $\piK$ and $\Kpi$ reactions. These reactions were 
studied using effective Lagrangian approaches~\cite{Hyodo:2003th,
Liu:2003rh,Oh:2003gj,Oh:2003kw}. Experiments for $\piK$ have been 
performed at KEK~\cite{ImaiMiwa}, and a high-resolution experiment 
for the $\Kpi$ reaction is ongoing. We can compare the results with
these experiments.

This paper is organized as follows. In the next section, we show the
framework of representation mixing and relevant experimental information
of nucleon decay. In Sec.~\ref{sec:Lag}, the effective interaction
Lagrangians for nucleons and for the antidecuplet are introduced for 
both $J^P=1/2^+$ and $3/2^-$ cases. The coupling constants are determined
in Sec.~\ref{sec:coupling} by considering the decay widths of $N^*$ 
resonances and the self-energy of the $\Theta^+$. With the effective
Lagrangians, the reaction processes $\piK$ and $\Kpi$ are analyzed in
Sec.~\ref{sec:reaction}. The final section is devoted to a summary.

\section{Representation mixing scheme 
and experimental information}\label{sec:mixing}

Let us briefly review the previous study of representation
mixing~\cite{Hyodo:2005wa} and summarize the experimental decays of
nucleon resonances. We have performed a phenomenological analysis on
the exotic particles using flavor SU(3) symmetry. It is found that the
masses of $\Theta(1540)$ and $\Xi_{3/2}(1860)$ are well fitted in an
antidecuplet ($\bm{\overline{10}}$) representation which mixes with
an octet ($\bm{8}$), with known baryon resonances of $J^P=1/2^+$ or
$3/2^-$. The $1/2^-$ case gives too large a decay width for the 
$\Theta^+$, and not enough resonances are well established for $3/2^+$
to complete the analysis. Under the representation mixing, the physical 
nucleon states are defined as
\begin{equation}
    \begin{split}
	 \ket{N_1} =& \ket{\bm{8},N} \cos\theta_N
	 - \ket{\tenbar,N} \sin\theta_N  , \\
	 \ket{N_2} =& \ket{\tenbar,N} \cos\theta_N
	 + \ket{\bm{8},N} \sin\theta_N  .
    \end{split}
    \label{eq:Nmixing}
\end{equation}
Two states $N_1$ and $N_2$ represents $N(1440)$ and $N(1710)$ for the 
$1/2^+$ case and $N(1520)$ and $N(1700)$ for the $3/2^-$ case. The 
mixing angles $\theta_N$ can be determined by experimental spectra of
known resonances as
\begin{align}
    \theta_N =& 29^{\circ}\quad
    \text{for } J^P=1/2^+ \ ,  \label{eq:thetaN1o2}  \\
    \theta_N =& 33^{\circ}\quad
    \text{for } J^P=3/2^- \ .
    \label{eq:thetaN3o2}
\end{align}
Both angles are close to the ideal mixing $\theta_N\sim 35.2^{\circ}$,
in which the nucleon states are classified by the number of strange
quarks (antiquarks). In other words, states are well mixed and the
effect of mixing of states is important.

\begin{figure}[tbp]
    \centering
    \includegraphics[width=8cm,clip]{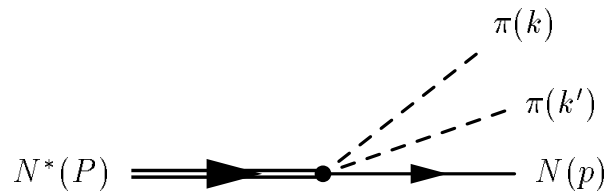}
    \caption{\label{fig:sdecay}
    Feynman diagram for the three-body decay of the $N^*$ resonance.}
\end{figure}

Using these mixing angles and decay widths of nucleon resonances
($\Gamma_{N^*\to\pi N}$), we can calculate the decay width of $\Theta$
($\Gamma_{\Theta\to KN}$) through the SU(3) relation between the 
coupling constants
\begin{equation}
    g_{\Theta}
    = \sqrt{6}(g_{N_2}\cos\theta_{N}-g_{N_1}\sin\theta_N),
    \nonumber
\end{equation}
where $g_{\Theta}$, $g_{N_1}$, and  $g_{N_2}$ are the coupling constants
of $\Theta$ and nucleon resonances. With the known coupling constants
$g_{N_1}$ and $g_{N_2}$, we obtained $\Gamma_{\Theta}\sim 30$ MeV for 
$J^P=1/2^+$ and $\Gamma_{\Theta}\sim 3$ MeV for $J^P=3/2^-$. Here we 
extend this approach to the three-body decay, as shown in 
Fig.~\ref{fig:sdecay}.

In Table~\ref{tbl:exp}, we show the experimental information of the
decay pattern of the nucleon resonances $N^*\to \pi\pi N$ taken from
the Particle Data Group (PDG)~\cite{Eidelman:2004wy}. For convenience,
we refer to $\pi\pi(I=0, s\text{ wave}) N$ and 
$\pi\pi(I=1, p\text{ wave}) N$ modes as ``scalar'' ($s$) and ``vector''
($v$), respectively. There is no information for the scalar decay of 
$N(1700)$. PDG only shows the fraction decaying via the $\pi\pi N$ mode
(85-95\%) and an upper bound for $\rho N$ mode ($<$35\%), although 
several intermediate states including $\pi\pi(I=0, s\text{ wave}) N$ 
are shown in the table. For the estimation of the coupling constants,
we adopt the total branching ratio to $\pi\pi N$ channel as the upper 
limit of the branch for  $\pi\pi (I=0)N$ state,
$BR_{N(1700)\to \pi\pi(I=0)N}<85\text{-}95\% $.

\begin{table}[btp]
    \centering
    \caption{Experimental information of two-pion decay of nucleon
    resonances. ``Scalar'' represents the mode 
    $\pi\pi(I=0, s\text{ wave}) N$ and ``Vector'' means 
    $\pi\pi(I=1, p\text{ wave}) N$ mode. Values in parenthesis are 
    averaged over the interval quoted in PDG~\cite{Eidelman:2004wy}.}
    \begin{ruledtabular}
    \begin{tabular}{ccccc}
	 $J^P$ & State & $\Gamma_{tot}$ [MeV]
	 & Scalar [\%]
	 & Vector [\%]	 \\
	\hline
	$1/2^+$ & N(1440)
	& 350
	& 5-10(7.5) & $<$8  \\
	& N(1710)
	& 100 
	& 10-40(25)  & 5-25(15) \\
	$3/2^-$ & N(1520) 
	& 120
	& 10-40(25)  & 15-25(20) \\
	& N(1700) 
	& 100 
	& $<85$-$95$\footnote{The scalar decay of $N(1700)$ is taken 
	from the total branching ratio to the $\pi\pi N$ channel.} 
	& $<$35 \\
    \end{tabular}
    \end{ruledtabular}
    \label{tbl:exp}
\end{table}

\section{Effective interaction Lagrangians}\label{sec:Lag}

Here we write down the effective Lagrangians that account for the 
interactions in the present analysis. We need two steps, namely, 
the extraction of the $\tenbar$ component from the $N^*\to \pi\pi N$
decay and the extrapolation to the $\Theta \pi K N$ channel. 
Lagrangians for nucleons will be used for the former purpose; the
Lagrangians for the antidecuplet will tell us the SU(3) relation
between channels in the multiplet.

In general, for an $N^*\to \pi\pi N$ vertex with an $N^*$ in octet 
or antidecuplet representations, there are several structures of 
interaction Lagrangians that are SU(3) symmetric. However, for octet
$N^*$, information of other channels are not relevant here, because
we do not want to study other channels. Therefore, we write down only
the $N^*\pi\pi N$ channels, instead of writing down all possible 
Lagrangians.

Using the partial decay widths of the two nucleon resonances 
$\Gamma^{s,v}_{i}$, we determine the absolute values of the coupling 
constants $|g_{i}^{s,v}|$, where superscripts $s$ and $v$ stand for 
the scalar- and vector-type correlations of two mesons. From them,
we can obtain the antidecuplet and octet components of the coupling
constants as
\begin{equation}
    \begin{split}
	g^{s,v}(\overline{10})
	&=-|g_1^{s,v}|\sin\theta_N\pm |g_2^{s,v}|\cos\theta_N ,  \\
	g^{s,v}(8)
	&=|g_1^{s,v}|\cos\theta_N\pm |g_2^{s,v}|\sin\theta_N ,
    \end{split}
    \label{eq:couplingmix}
\end{equation}
based on Eq.~\eqref{eq:Nmixing}. Since the relative phase of the two
coupling constants cannot be determined, the $\pm$ sign appears.
Here we use $\theta_N$ obtained from the mass spectra as shown in
Eqs.~\eqref{eq:thetaN1o2} and \eqref{eq:thetaN3o2}. When the coupling
constants have experimental uncertainties, we vary them within the region
and check the minimum and maximum of corresponding values of 
$g^{s,v}(\overline{10})$.

\subsection{Lagrangians for nucleons with $J^P=1/2^+$}

Let us consider the $J^P=1/2^+$ case. The interaction Lagrangians for
nucleons can be written as
\begin{equation}
    \mathcal{L}^{s}_i
    =\frac{g^{s}_i}{2\sqrt{2}f}
    \overline{N}^*_i\bm{\pi}\cdot \bm{
    \pi} N+ \text{h.c.}
    \label{eq:sLag1o2}
\end{equation}
and
\begin{align}
    \mathcal{L}^{v}_i
    &=i\frac{g^{v}_i}{4\sqrt{2}f^2}
    \overline{N}^*_i
    (\bm{\pi}\cdot\overleftrightarrow{\Slash{\partial}} \bm{\pi}) 
    N+ \text{h.c.} \nonumber \\
    &=i\frac{g^{v}_i}{4\sqrt{2}f^2}
    \overline{N}^*_i
    (\bm{\pi}\cdot\Slash{\partial}\bm{\pi}
    -\Slash{\partial}\bm{\pi}\cdot\bm{\pi}) 
    N+ \text{h.c.} ,
    \label{eq:vLag1o2}
\end{align}
where $f=93$ MeV is the pion decay constant, $g^{s,v}_i$ are
dimensionless coupling constants, and h.c. stands for the hermitian
conjugate. Subscript $i=1,2$ denotes the two nucleons $N(1440)$ and
$N(1710)$, respectively. The numerical factors are chosen such that
the coupling constants $g^{s,v}_i$ should be consistent with the
Lagrangians for the antidecuplet, which will be given later. For 
nucleon, $N^*$, and pion fields, we adopt the convention
\begin{equation}
    N=
    \begin{pmatrix}
	p \\
	n
    \end{pmatrix},
    \quad
    N^*_i=
    \begin{pmatrix}
	p^*_i \\
	n^*_i
    \end{pmatrix},
    \quad
    \bm{\pi}=
    \begin{pmatrix}
	\pi^0 & \sqrt{2}\pi^+ \\
	\sqrt{2}\pi^- & -\pi^0
    \end{pmatrix}
    \label{eq:fielddef} .
\end{equation}

\subsection{Lagrangians for the antidecuplet with $1/2^+$}

To connect the coupling constant of the process $N^*\pi\pi N$ to that 
of $\Theta K\pi N$, we write down the interaction Lagrangian for the 
antidecuplet. Flavor SU(3) structure of these terms are studied in
Ref.~\cite{Hosaka:2004mv}. In the present case, for the scalar-type 
correlation, we have
\begin{equation}
    \mathcal{L}^{s}_{1/2^+}
    =\frac{g^{s}_{1/2^+}}{2f}
    \overline{P}_{ijk}\epsilon^{lmk}
    \phi_{l}{}^{a}\phi_{a}{}^{i} B_{m}{}^{j}
    + \text{h.c.} ,
    \label{eq:8sLag1o2}
\end{equation}
whereas for the vector-type correlation, we have
\begin{equation}
    \mathcal{L}^{v}_{1/2^+}
    =i\frac{g^{v}_{1/2^+}}{4f^2}
    \overline{P}_{ijk}\epsilon^{lmk}\gamma^{\mu}
    (\partial_{\mu}\phi_{l}{}^{a} \phi_{a}{}^{i}
    - \phi_{l}{}^{a}\partial_{\mu} \phi_{a}{}^{i})B_{m}{}^{j}
    + \text{h.c.}
    \label{eq:8aLag1o2}
\end{equation}
In Eqs.~\eqref{eq:8sLag1o2} and \eqref{eq:8aLag1o2}, the coupling
constants are for the antidecuplet baryon, which corresponds to 
Eq.~\eqref{eq:couplingmix}. These Lagrangians correspond to 
$\mathcal{L}^{8s}$ and $\mathcal{L}^{8a}$ in Ref.~\cite{Hosaka:2004mv}.
The octet meson (baryon) field $\phi$ ($B$) and the antidecuplet baryon
field $P$ are defined as
\begin{align}
    \phi=&
    \begin{pmatrix}
	\frac{1}{\sqrt{2}}\pi^{0}+\frac{1}{\sqrt{6}}\eta & 
	\pi^{+} & K^{+} \\ 
	\pi^{-} & -\frac{1}{\sqrt{2}}\pi^{0}
	+\frac{1}{\sqrt{6}}\eta & K^{0} \\ 
	K^{-} & \overline{K}^{0} & -\frac{2}{\sqrt{6}}\eta
    \end{pmatrix} ,
    \label{eq:mesonLag} \\
    B=&
    \begin{pmatrix}
	\frac{1}{\sqrt{2}}\Sigma^{0}+\frac{1}{\sqrt{6}}\Lambda & 
	\Sigma^{+} & p \\ 
	\Sigma^{-} & -\frac{1}{\sqrt{2}}\Sigma^{0}
	+\frac{1}{\sqrt{6}}\Lambda & n \\ 
	\Xi^{-} & \Xi^{0} & -\frac{2}{\sqrt{6}}\Lambda
    \end{pmatrix} ,
    \label{eq:baryonLag} \\
    P^{333} &= \sqrt{6}\Theta^{+}_{\overline{10}}  , 
    \quad
    P^{133} = \sqrt{2} \, N^0_{\overline{10}}  ,  \nonumber \\*
    P^{233} &= -\sqrt{2} \, N^+_{\overline{10}}  ,
    \quad
    P^{113} = \sqrt{2} \, \Sigma^{-}_{\overline{10}} ,\nonumber \\*
    P^{123} &= -\Sigma^{0}_{\overline{10}} ,
    \quad
    P^{223} = -\sqrt{2} \, \Sigma^{+}_{\overline{10}}  , 
    \label{eq:pentaLag} \\*
    P^{111} &=  \sqrt{6}\Xi^{--}_{\overline{10}} , 
    \quad
    P^{112} = -\sqrt{2}  \, \Xi_{\overline{10}}^{-}  , \nonumber \\*
    P^{122} &= \sqrt{2} \, \Xi_{\overline{10}}^{0} ,
    \quad
    P^{222} = - \sqrt{6}\Xi_{\overline{10}}^{+}  .\nonumber
\end{align}
Note that 
the coefficients for $N^*\pi\pi N$ in the expansion of the Lagrangians
are the same as Eqs.~\eqref{eq:sLag1o2} and \eqref{eq:vLag1o2}, 
respectively. This means that the normalization of the coupling 
constants in both Lagrangians are the same.

There is another Lagrangian for the scalar-type correlation 
$\mathcal{L}^{27}$~\cite{Hosaka:2004mv}. However, the contribution of 
this term can be expressed by the following parametrization:
\begin{equation}
    a\mathcal{L}^{8s}+b\mathcal{L}^{27}  ,
    \quad
    b=-\tfrac{5}{4}(1-a) ,
    \label{eq:Lagcomb1}
\end{equation}
with $g^{8s}=g^{27}$. The ratio of $\mathcal{L}^{8s}$ and 
$\mathcal{L}^{27}$ is controlled by the parameter $a$, without changing
the total coupling constant of $N^* \pi\pi N$. The important point is 
that this combination of the two Lagrangians also does not change the 
$\Theta K\pi N$ channel, as we can see in the table in 
Ref.~\cite{Hosaka:2004mv}. Therefore, in the present purpose, it is
sufficient to consider the Lagrangians \eqref{eq:8sLag1o2} and 
\eqref{eq:8aLag1o2}.

\subsection{Lagrangians for nucleons with $J^P=3/2^-$}

We express the spin $3/2$ baryons as Rarita-Schwinger fields
$B^{\mu}$~\cite{Rarita:1941mf}. The effective Lagrangians can be 
written as
\begin{align}
    \mathcal{L}^{s}_{i}
    &=i\frac{g^{s}_{i}}{4\sqrt{2}f^2}
    \overline{N}_i^{*\mu}\partial_{\mu}(\bm{\pi}\cdot \bm{
    \pi}) N+ \text{h.c.} \nonumber \\
    &=i\frac{g^{s}_{i}}{4\sqrt{2}f^2}
    \overline{N}_i^{*\mu}
    (\partial_{\mu}\bm{\pi}\cdot \bm{\pi}
    +\bm{\pi}\cdot \partial_{
    \mu} \bm{\pi}) N+ \text{h.c.} 
    \label{eq:sLag3o2}
\end{align}
and
\begin{equation}
    \mathcal{L}^{v}_{i}
    =i\frac{g^{v}_{i}}{4\sqrt{2}f^2}
    \overline{N}_{i}^{*\mu}(\bm{\pi}\cdot 
    \overleftrightarrow{\partial_{\mu}} 
    \bm{\pi}) N+ \text{h.c.} 
    \label{eq:vLag3o2}
\end{equation}
Here $i=1,2$ denotes the two nucleons $N(1520)$ and $N(1700)$,
respectively. Notice that a derivative of meson field is needed for the
scalar Lagrangian whose Dirac index is to be contracted with that of
Rarita-Schwinger field. Since the flavor structure of these Lagrangians
is the same as in Eqs.~\eqref{eq:sLag1o2} and \eqref{eq:vLag1o2},
we will have the same coefficients up to the coupling factors. The 
antidecuplet component of the coupling constants can be determined as 
in Eq.~\eqref{eq:couplingmix}.

\subsection{Lagrangians for the antidecuplet with $3/2^-$}

We write the Lagrangians for the antidecuplet as a straightforward 
extension of those in $1/2^+$ case:
\begin{equation}
    \mathcal{L}^{s}_{3/2^-}
    =i\frac{g^{s}_{3/2^-}}{4f^2}
    \overline{P}_{ijk}^{\mu}\epsilon^{lmk}
    \partial_{\mu}(\phi_{l}{}^{a}\phi_{a}{}^{i})
    B_{m}{}^{j}+ \text{h.c.}
    \label{eq:8sLag3o2}
\end{equation}
while for the vector type correlation we have
\begin{equation}
    \mathcal{L}^{v}_{3/2^-}
    =i\frac{g^{v}_{3/2^-}}{4f^2}
    \overline{P}_{ijk}^{\mu}\epsilon^{lmk}
    (\partial_{\mu}\phi_{l}{}^{a} \phi_{a}{}^{i}
    - \phi_{l}{}^{a}\partial_{\mu} \phi_{a}{}^{i})B_{m}{}^{j}
    + \text{h.c.}
    \label{eq:8aLag3o2}
\end{equation}
Here the flavor structure is the same as in Eq.~\eqref{eq:8aLag1o2}.

\section{Numerical results for the coupling 
constants}\label{sec:coupling}

To study the coupling constants, let us start with the decay width of 
a resonance into two mesons and one baryon, which is given by
\begin{align}
    \Gamma_{\text{BMM}}
    =&\int \frac{d^3p}{(2\pi)^3}
    \frac{M}{E(p)}
    \int \frac{d^3k}{(2\pi)^3}\frac{1}{2\omega(k)}
    \int \frac{d^3k^{\prime}}{(2\pi)^3}
    \frac{1}{2\omega^{\prime}(k^{\prime})}
    \nonumber \\
    &\times 
    \overline{\Sigma}\Sigma|t(\omega,\omega^{\prime},\cos\theta)|^2
    (2\pi)^4
    \delta^{(4)}(P-p-k-k^{\prime})
    \nonumber \\
    =&\frac{M}{16\pi^3}
    \int_{\omega_{\text{min}}}^{\omega_{\text{max}}} d\omega 
    \int_{\omega^{\prime}_{\text{min}}}^{\omega^{\prime}_{\text{max}}} 
    d\omega^{\prime}
    \nonumber \\
    &\times \overline{\Sigma}\Sigma|t(\omega,\omega^{\prime},a)|^2
    \Theta(1-a^2),
    \label{eq:decaywidth}
\end{align}
with
\begin{equation}
    \begin{split}
	\omega_{\text{min}} 
	=& m ,\\
	\omega_{\text{max}}
	=& \frac{M_R^2-M^2-2Mm}{2M_R} , \\
	a
	=& \frac{(M_R-\omega-\omega^{\prime})^2
	-M^2-|\bm{k}|^2-|\bm{k}^{\prime}|^2}{2|\bm{k}||\bm{k}^{\prime}|},
    \end{split}
    \nonumber
\end{equation}
where we assign the momentum variables $P=(M_R,\bm{0})$, 
$k=(\omega,\bm{k})$, $k^{\prime}=(\omega^{\prime},\bm{k}^{\prime})$, and
$p=(E,\bm{p})$ as in Fig.~\ref{fig:sdecay}; $M_R$, $M$, and $m$ are the 
masses of the resonance, baryon, and mesons, respectively; and $\theta$
is the angle between the momenta $\bm{k}$ and $\bm{k}^{\prime}$. The
on-shell energies of particles are given by $\omega=\sqrt{m^2+\bm{k}^2}$, 
$\omega^{\prime}=\sqrt{m^2+(\bm{k}^{\prime})^2}$, and
$E=\sqrt{M^2+\bm{p}^2}$; $\Theta$ denotes the step function; and 
$\overline{\Sigma}\Sigma$ stands for the spin sum of the fermion states.

In the following, we evaluate the squared amplitude 
$\overline{\Sigma}\Sigma|t(\omega,\omega^{\prime},\cos\theta)|^2$
for the $N^*\to \pi\pi N$ decay in the nonrelativistic approximation.
For the $1/2^+$ case, from Eq.~\eqref{eq:sLag1o2}, the scalar Lagrangian
gives the term
\begin{equation}
    \begin{split}
	\overline{\Sigma}\Sigma|t_{1/2^+}^{s}|^2
	&= 3\left(\frac{g^{s}_{1/2^+}}{2f}\right)^2\frac{E+M}{2M}.
    \end{split}
    \label{eq:samp1o2}
\end{equation}
Note that we include the normalization factor $(E+M)/2M$ to be 
consistent with the other amplitude, although the effect of this factor
is small (of the order of a few percent) in the results.

\begin{figure}[tbp]
    \centering
    \includegraphics[width=8cm,clip]{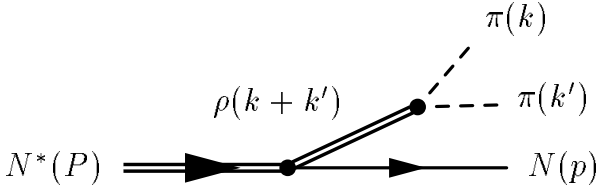}
    \caption{\label{fig:vdecay}
    Three-body decay of the $N^*$ resonance
    with insertion of the vector meson propagator.}
\end{figure}

For the vector-type coupling, we insert the vector meson propagator
to account for the $\rho$ meson correlation~\cite{Hosaka:2004mv},
as shown in Fig.~\ref{fig:vdecay}. Then the squared amplitude becomes
\begin{align}
    \overline{\Sigma}\Sigma|t^v_{1/2^+}|^2 
    =&
    6\left(\frac{g^{v}_{1/2^+}}{4f^2}\right)^2
    \frac{1}{2M}\Bigl\{
    (E+M)(\omega-\omega^{\prime})^2 \nonumber \\
    &+2(|\bm{k}|^2-|\bm{k}^{\prime}|^2)(\omega-\omega^{\prime})
    \nonumber \\
    &+(E-M)(\bm{k}-\bm{k}^{\prime})^2
    \Bigr\} \nonumber \\
    &\times \left|\frac{-m_{\rho}^2}{s^{\prime}-m_{\rho}^2
    +im_{\rho}\Gamma(s^{\prime})}\right|^2 ,
    \label{eq:vamp1o2}
\end{align}
where $m_{\rho}$ is the mass of $\rho$ meson, 
$s^{\prime}=(k+k^{\prime})^2$, and $\Gamma(s^{\prime})$ is the 
energy-dependent width given by
\begin{equation}
    \Gamma(s^{\prime})
    =\Gamma_{\rho} \times \left(
    \frac{p_{\text{cm}}(s^{\prime})}{p_{\text{cm}}(m_{\rho}^2)}
    \right)^3 ,
    \nonumber
\end{equation}
with the three-momentum of the final particles in the $\rho$ rest frame
given by
\begin{equation}
    p_{\text{cm}}(s^{\prime}) =
    \begin{cases}
	\frac{\lambda^{1/2}(s^{\prime},m_{\pi}^2,m_{\pi}^2)}
	{2\sqrt{s^{\prime}}}
	& \text{for } s^{\prime}> 4m_{\pi}^2 ,\\
	0 & \text{for } s^{\prime}\leqslant 4m_{\pi}^2 ,
    \end{cases}
    \nonumber
\end{equation}
and $\lambda(a,b,c)$ is the K\"allen function. Note that in 
Eq.~\eqref{eq:vamp1o2} we take the terms up to next to leading order 
in the nonrelativistic expansion, since the leading order term
$(\omega-\omega^{\prime})$ appears as the difference of two energies,
which can be zero.

The squared amplitudes for $J^P=3/2^-$ can be obtained in a similar way:
\begin{align}
    \overline{\Sigma}\Sigma|t^{s}_{3/2^-}|^2
    =&\left(\frac{g_{3/2^-}^{s}}{4f^2}\right)^2
    (\bm{k}+\bm{k}^{\prime})^2\frac{E+M}{2M}
    \label{eq:samp3o2} ,\\
    \overline{\Sigma}\Sigma|t^{v}_{3/2^-}|^2
    =&
    2
    \left(\frac{g_{3/2^-}^{v}}{4f^2}\right)^2
    (\bm{k}-\bm{k}^{\prime})^2\frac{E+M}{2M} \nonumber \\
    &\times \left|\frac{-m_{\rho}^2}{s^{\prime}-m_{\rho}^2
    +im_{\rho}\Gamma(s^{\prime})}\right|^2.
    \label{eq:vamp3o2}
\end{align}

\subsection{Numerical result for the $J^P=1/2^+$ case}

Now we evaluate the coupling constants numerically. Using the averaged
values in Table~\ref{tbl:exp}, we obtain the coupling constants $g^s_i$
and $g^v_i$ for these channels:
\begin{align}
    |g^s_{N(1440)}|
    &= 4.28, 
    &|g^v_{N(1440)}|
    &<3.68, \label{eq:1440res} \\
    |g^s_{N(1710)}|
    &= 1.84, 
    &|g^v_{N(1710)}|
    &=0.31. \label{eq:1710res}
\end{align}
By substituting them into Eq.~\eqref{eq:couplingmix} (but suppressing 
the label $\overline{\bm{10}}$ for simplicity), the antidecuplet
components are extracted as
\begin{align}
    |g^{s}_{1/2^+}|
    &= 0.47 ,\quad 3.68 ,
    \label{eq:1o2res}
\end{align}
where two values correspond to the results with different relative 
phases between the two coupling constants. For $|g^{v}_{1/2^+}|$, 
only the upper bound is given for $N(1440)$; therefore we can not fix 
the central value.

\begin{figure}[tbp]
    \centering
    \includegraphics[width=8cm,clip]{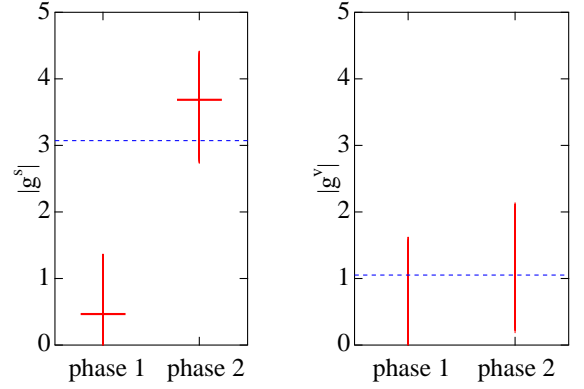}
    \caption{\label{fig:coupling1o2}
    (Color online) Numerical results for the coupling constants
    with $J^P=1/2^+$. The two choices of the relative phase between 
    coupling constants are marked as ``phase 1'' and ``phase 2''.
    Allowed regions of the coupling constants are shown by the vertical 
    bar. Horizontal bars represent the results obtained with the 
    averaged values, which are absent for the vector case. Horizontal
    dashed lines show the upper limits of the coupling constants derived
    from the self-energy $|\re\Sigma|<200$ MeV.}
\end{figure}

When we take into account the experimental uncertainties in branching
ratio, the antidecuplet components can vary within the following ranges:
\begin{align*}
    0
    &< |g^{s}_{1/2^+}| < 1.37,
    & 0
    &< |g^{v}_{1/2^+}|< 2.14,
    \\
    2.72
    &< |g^{s}_{1/2^+}| < 4.42,
\end{align*}
including both cases for the phase. These uncertainties are also shown
by the vertical bar in Fig.~\ref{fig:coupling1o2}, with the horizontal
bars being the result with the averaged value in Eq.~\eqref{eq:1o2res}.

Now let us consider the phenomenological implication of this result.
In the study of self-energy~\cite{Hosaka:2004mv}, the coupling constants 
have been derived by assuming that the $\Theta^+$ belongs to a pure 
antidecuplet together with $N(1710)$, where we have determined 
$|g^{s}_{1/2^+}|=1.88$ and $|g^{v}_{1/2^+}|=0.315$ [essentially the same
as values in Eq.~\eqref{eq:1710res}]. In the calculation of the 
self-energy of $\Theta^+$, the effect of the mixing only changes the 
coupling constants, by neglecting the small contribution from
$\mathcal{L}^{27}$. In this case, the $\Theta^+$ self-energy with the
new coupling constants can be written as
\begin{align}
    \Sigma^s_{\Theta^+}(g^{s}_{1/2^+})
    =&\Sigma^s_{\Theta^+}(1.88)\times
    \frac{|g^{s}_{1/2^+}|^2}{1.88^2} ,
    \label{eq:thetas} \\
    \Sigma^v_{\Theta^+}(g^{v}_{1/2^+})
    =&\Sigma^v_{\Theta^+}(0.315)\times
    \frac{|g^{v}_{1/2^+}|^2}{0.315^2} . 
    \label{eq:thetav}
\end{align}
The real parts of the self-energy depend on the initial energy and the
cutoff value of the loop integral. We have estimated
$\re\Sigma^s_{\Theta^+}(g=1.88)\sim -75$ MeV and
$\re\Sigma^v_{\Theta^+}(g=0.315))\sim -18$ MeV for an initial energy of
$1540$-$1700$ MeV and with a cutoff of $700$-$800$ MeV. Using 
Eqs.~\eqref{eq:thetas} and \eqref{eq:thetav} with the values of 
Eq.~\eqref{eq:1o2res}, we obtain
\begin{align}
    \Sigma^s_{\Theta^+}
    &= -287, \quad -4.7 \text{ MeV} ,
    &0>\Sigma^v_{\Theta^+}
    &> -770\text{ MeV} .
\end{align}
The sum of these values are the contribution to the self-energy of 
$\Theta^+$ from the two-meson cloud. Naively, we expect that it should 
be of the order of 100 MeV, at most $\sim$20\% of the total 
energy~\cite{Theberge:1980ye,Hosaka:2004mv}. From this consideration,
we adopt the condition that the magnitude of one of the contributions
should not exceed $200$ MeV: $|\re\Sigma^v_{\Theta^+}|<200$.

For the scalar coupling, this condition is satisfied when
\begin{equation}
    |g^{s}_{1/2^+}|<3.07.
    \label{eq:slimit}
\end{equation}
Therefore, we can exclude the choice of ``phase 2'' in 
Fig.~\ref{fig:coupling1o2}. In the same way, the upper limit of 
$|g^{v}_{1/2^+}|$ should be imposed as
\begin{equation}
    |g^{v}_{1/2^+}|<1.05
    \label{eq:vlimit}
\end{equation}
to be consistent with the condition $|\re\Sigma^v_{\Theta^+}|<200$ MeV.
This is compatible with Eq.~\eqref{eq:1o2res}, although 
Eq.~\eqref{eq:vlimit} gives a more stringent constraint. These upper
limits are also shown in Fig.~\ref{fig:coupling1o2} by the dashed lines.

\subsection{Numerical result for the $J^P=3/2^-$ case}

\begin{figure}[tbp]
    \centering
    \includegraphics[width=8cm,clip]{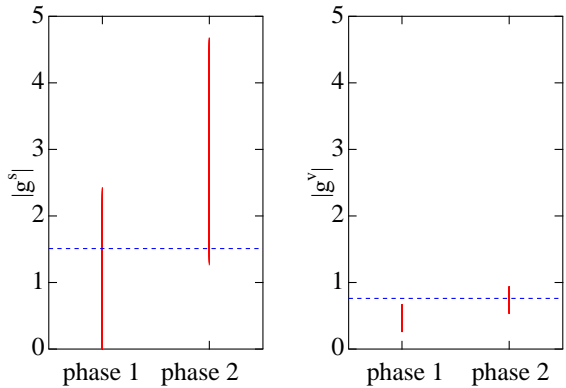}
    \caption{\label{fig:coupling3o2}
    (Color online) Numerical results for the coupling constants
    with $J^P=3/2^-$. The two choices of the relative phase between
    coupling constants are marked as ``phase 1'' and ``phase 2''.
    Allowed regions of the coupling constants are shown by the vertical 
    bar. Horizontal dashed lines show the upper limit of the coupling 
    constants derived from the self-energy $|\re\Sigma|<200$ MeV.}
\end{figure}

Now we consider the $J^P=3/2^-$ case. Using the central values in 
Table~\ref{tbl:exp}, we obtain the coupling constants $g^s_i$ and 
$g^v_i$ for these channels:
\begin{align}
    |g^s_{N(1520)}|
    &= 3.56, 
    &|g^v_{N(1520)}|
    &=1.11, \label{eq:1520res} \\
    |g^s_{N(1700)}|
    &< 2.66,
    &|g^v_{N(1700)}|
    &<0.32. \label{eq:1700res}
\end{align}
In this case, with the same reason as in the vector coupling for the 
$1/2^+$ case, the central value cannot be determined. Experimental
uncertainties allows the antidecuplet components to vary within the 
following ranges:
\begin{align}
    0
    &< |g^{s}_{3/2^-}| < 4.68,
    &0.25
    &< |g^{v}_{3/2^-}|< 0.94,
    \label{eq:range1_3o2}
\end{align}
including both cases for the phase. The result are shown by the vertical
bars in Fig.~\ref{fig:coupling3o2}.

It is worth noting that the region of $|g^{v}_{3/2^-}|$ does not reach
zero, even though the $|g^v_{N(1700)}|$ can be zero. The condition for
$g^{s,v}(\tenbar)=0$ leads to
\begin{equation}
    \frac{|g_2^{s,v}|}{|g_1^{s,v}|}
    =\tan\theta_N\sim
    \begin{cases}
	 0.55 & \text{for }1/2^+ ,  \\
	 0.65 & \text{for }3/2^- .
    \end{cases}
    \label{eq:zerocond}
\end{equation}
This means that $g^{s,v}(\tenbar)$ becomes zero only if the 
condition~\eqref{eq:zerocond} is satisfied within the uncertainty of
coupling constants.

We can also estimate the magnitude of the self-energy, by substituting 
the squared amplitudes for $3/2^-$ case in the formulas of the 
self-energy shown in Ref.~\cite{Hosaka:2004mv}. For 
$|g^{s}_{3/2^-}|=  4.17$, we estimate the real part of the self-energy 
as $-1518$ MeV for an initial energy of $1540$-$1700$ MeV and a cutoff 
of $700$-$800$ MeV. This huge self-energy for $3/2^-$ case is due to the 
$p$-wave nature of the two-meson coupling, namely, the existence of a
momentum variable in the loop integral. A similar large self-energy was
observed when the self-energy is calculated with the
chiral Lagrangian in Ref.~\cite{Hosaka:2004mv}.
Thus, to have some reasonable values
for the self-energy $|\re\Sigma^s_{\Theta^+}|<200$ MeV,
\begin{equation}
    |g^{s}_{3/2^-}|<1.51 .
    \label{eq:slimit3o2}
\end{equation}

In the same way, for the vector term with $|g^{v}_{3/2^-}|=  0.61$,
we estimate the real part of the self-energy as $-130$ MeV. In this 
case, the self-energy is suppressed by the vector meson propagator.
The use of small number 0.61 for the coupling constant also accounts
for the small value of the self-energy. The condition of the self-energy
$|\re\Sigma^v_{\Theta^+}|<200$ MeV gives the constraints
\begin{equation}
    |g^{v}_{3/2^-}|<0.76 .
    \label{eq:vlimit3o2}
\end{equation}
Both upper limits \eqref{eq:slimit3o2} and \eqref{eq:vlimit3o2} are 
indicated by horizontal dashed lines in Fig.~\ref{fig:coupling3o2}.

\section{Analysis of the meson-induced reactions}\label{sec:reaction}

\begin{figure}[tbp]
    \centering
    \includegraphics[width=8cm,clip]{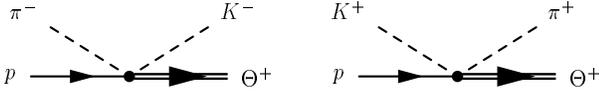}
    \caption{\label{fig:reaction}
    Feynman diagrams for the meson-induced reactions
    for $\Theta^+$ production.}
\end{figure}

\begin{figure}[tbp]
    \centering
    \includegraphics[width=5cm,clip]{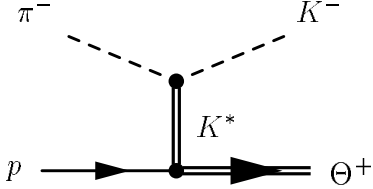}
    \caption{\label{fig:reactionvec}
    Feynman diagram for the 
    meson-induced reaction for $\Theta^+$ production
    with a vector meson propagator.}
\end{figure}

As an application of effective Lagrangians, we calculate the reaction
processes $\piK$ and $\Kpi$ via tree-level diagrams as shown in 
Fig.~\ref{fig:reaction}. These are alternative reactions to, for 
instance, photo-induced reactions, which are useful for further study of
the $\Theta^+$. The amplitudes for these reactions are given by
\begin{align}
    -it_{1/2^+}^{s}&(\piK)
    =-it_{1/2^+}^{s}(\Kpi)\nonumber \\
    =&i\frac{g_{1/2^+}^{s}}{2f}(-\sqrt{6})
    N_{\Theta^+}N_p , 
    \label{eq:samp1o2reac} \\
    -it_{1/2^+}^{v}&(\piK)
    =it_{1/2^+}^{v}(\Kpi)\nonumber \\
    =& i\frac{g_{1/2^+}^{v}}{4f^2}(-\sqrt{6})
    (2\sqrt{s}-M_{\Theta}-M_p) 
    N_{\Theta^+}N_p
    F(k-k^{\prime}) 
    \label{eq:vamp1o2reac}
\end{align}
for the $1/2^+$ case and by
\begin{align}
    -it_{3/2^-}^{s}&(\piK)
    =-it_{3/2^-}^{s}(\Kpi)\nonumber \\
    =&i\frac{g_{3/2^-}^{s}}{4f^2}(-\sqrt{6})
    (\bm{k}-\bm{k}^{\prime})\cdot \bm{S}
    N_{\Theta^+}N_p , 
    \label{eq:samp3o2reac} \\
    -it_{3/2^-}^{v}&(\piK)
    =it_{3/2^-}^{v}(\Kpi)\nonumber \\
    =& -i\frac{g_{3/2^-}^{v}}{4f^2}(-\sqrt{6})
    (\bm{k}+\bm{k}^{\prime})\cdot \bm{S}
    N_{\Theta^+}N_p
    F(k-k^{\prime})
    \label{eq:vamp3o2reac}
\end{align}
for the $3/2^-$ case, where the normalization factor is 
$N_i=\sqrt{(E_i+M_i)/2M_i}$, $\bm{S}$ is the spin transition operator,
$\sqrt{s}$ is the initial energy, and $k$ and $k^{\prime}$ are the 
momenta of the incoming and outgoing mesons, respectively. Here we 
define the vector meson propagator (Fig.~\ref{fig:reactionvec}) as
\begin{equation}
    F(k-k^{\prime})=\frac{-m_{K^*}^2}{(k-k^{\prime})^2-m_{K^*}^2
    +im_{K^*}\Gamma[(k-k^{\prime})^2]},
    \label{eq:formfactor}
\end{equation}
which is included in the vector-type amplitude. In the kinematical
region in which we are interested, the momentum-dependent decay width
of $K^*$, $\Gamma((k-k^{\prime})^2)$ vanishes. Note that the scalar-type
amplitude gives the same sign for $\piK$ and $\Kpi$, whereas the vector
one gives opposite signs, reflecting the symmetry under exchange of two
meson fields in the effective Lagrangians.

Since the two amplitudes must be summed coherently, the squared 
amplitudes are given by
\begin{align}
    \overline{\Sigma}\Sigma|t_{1/2^+}|^2 
    = &\overline{\Sigma}\Sigma|t^s_{1/2^+}\pm t^v_{1/2^+}|^2 
    \nonumber  \\
    = &
    6\left(\frac{1}{2f}\right)^2 N_{\Theta^+}^2N_p^2
    \Bigl[(g_{1/2^+}^{s})^2
    \pm 2g_{1/2^+}^{s}g_{1/2^+}^{v} \nonumber \\
    &\times
    \frac{2\sqrt{s}-M_{\Theta}-M_p}{2f}F(k-k^{\prime}) 
    + (g_{1/2^+}^{v})^2
    \nonumber \\
    & \times
    \frac{(2\sqrt{s}-M_{\Theta}-M_p)^2}{4f^2}F^2(k-k^{\prime}) 
    \Bigr] ,
    \label{eq:ampsqure1o2}  \\
    \overline{\Sigma}\Sigma|t_{3/2^-}|^2  
    = &4
    \left(\frac{1}{4f^2}\right)^2 
    N_{\Theta^+}^2N_p^2
    \Bigl[
    (g_{3/2^-}^{s})^2 (\bm{k}-\bm{k}^{\prime})^2
    \nonumber \\
    &
    \mp 2g_{3/2^-}^{s}g_{3/2^-}^{v}(|\bm{k}|^2-|\bm{k}^{\prime}|^2)
    F(k-k^{\prime}) \nonumber \\
    &+(g_{3/2^-}^{v})^2(\bm{k}+\bm{k}^{\prime})^2F^2(k-k^{\prime})
    \Bigr] ,
    \label{eq:ampsqure3o2}
\end{align}
where $\pm$ and $\mp$ signs denote the $\piK$ and $\Kpi$ reactions,
respectively. Notice that the relative phase between the two coupling
constants is important, which affects the interference term of the two 
amplitudes. To determine the phase, we use the experimental information
from $\piK$ reaction at KEK~\cite{ImaiMiwa}, where the upper limit of 
the cross section has been extracted to be a few $\mu$b.

The differential cross section for these reactions is given by 
\begin{align}
    \frac{d\sigma}{d\cos\theta} (\sqrt{s},\cos\theta)  
    =&\frac{1}{4\pi s}\frac{|\bm{k}^{\prime}|}{|\bm{k}|}
    M_{p}M_{\Theta}
    \frac{1}{2}
    \overline{\Sigma}\Sigma|t(\sqrt{s},\cos\theta)|^2  ,
    \label{eq:dcross}
\end{align}
which is evaluated in the center-of-mass frame. The total cross section
can be obtained by integrating Eq.~\eqref{eq:dcross} with respect to
$\cos\theta$:
\begin{align}
    \sigma (\sqrt{s})  
    =& \int_{-1}^{1}d\cos\theta
    \frac{d\sigma}{d\cos\theta} (\sqrt{s},\cos\theta) .
    \nonumber
\end{align}

\subsection{Qualitative analysis for $J^P=1/2^+$ and $3/2^-$}

Now let us calculate the cross section using the coupling constants
obtained previously. In this section, we focus on the qualitative 
difference between $J^P=1/2^+$ and $3/2^-$ cases.  A more quantitative
estimation of cross sections will be given in later sections.

\begin{figure}[tbp]
    \centering
    \includegraphics[width=8cm,clip]{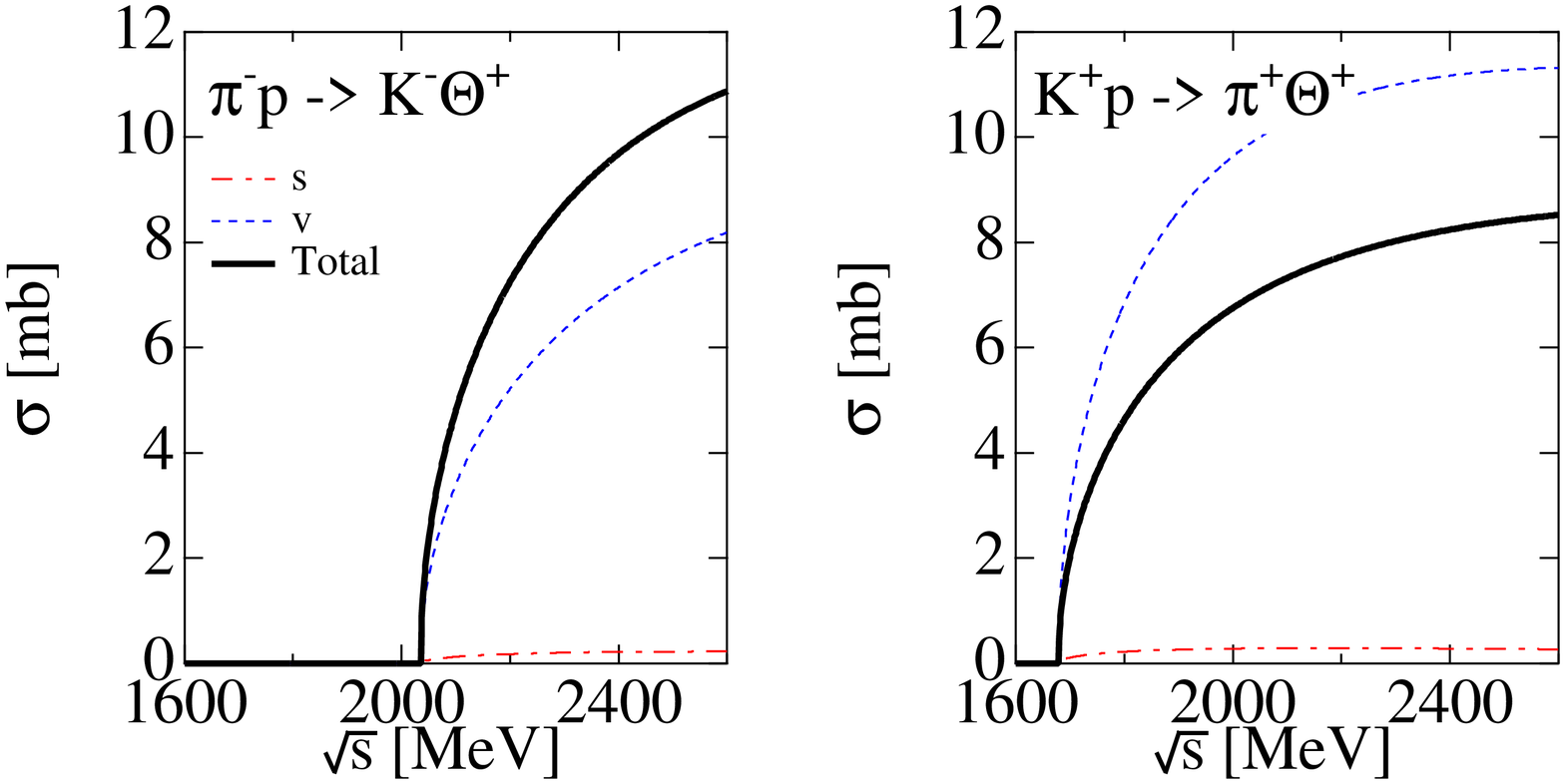}
    \caption{\label{fig:1o2_1}
    (Color online) Total cross sections
    for the $J^P=1/2^+$ case with $g^s=0.47$ and $g^v=0.47$.
    The thick line shows the result with full amplitude.
    Dash-dotted and dashed lines are the contributions from
    $s$ and $v$ terms, respectively.}
\end{figure}

We first calculate for the $1/2^+$ case, with coupling constants
\begin{equation}
    g^s_{1/2^+} = 0.47,\quad g^v_{1/2^+} = 0.47,
    \label{eq:1o2_1}
\end{equation}
where $g^s_{1/2^+}$ is one of the solutions that satisfies the 
condition~\eqref{eq:slimit}. Since the result~\eqref{eq:1o2res} spreads 
over a wide range, we choose $g^v_{1/2^+}=g^s_{1/2^+}$, which is well 
within the interval~\eqref{eq:vlimit} determined from the self-energy. 
The result is shown in Fig.~\ref{fig:1o2_1}, with contributions from 
$s$ and $v$ terms. Each contribution is calculated by switching off the
other term. As we see, the use of the same coupling constant for both 
terms results in the dominance of the vector term. However, there is a 
sizable interference effect between $s$ and $v$ terms, although the 
contribution from the $s$ term itself is small. The two amplitudes 
interfere constructively for the $\piK$ channel, whereas in the $\Kpi$
case they destructively interfere.

\begin{figure}[tbp]
    \centering
    \includegraphics[width=8cm,clip]{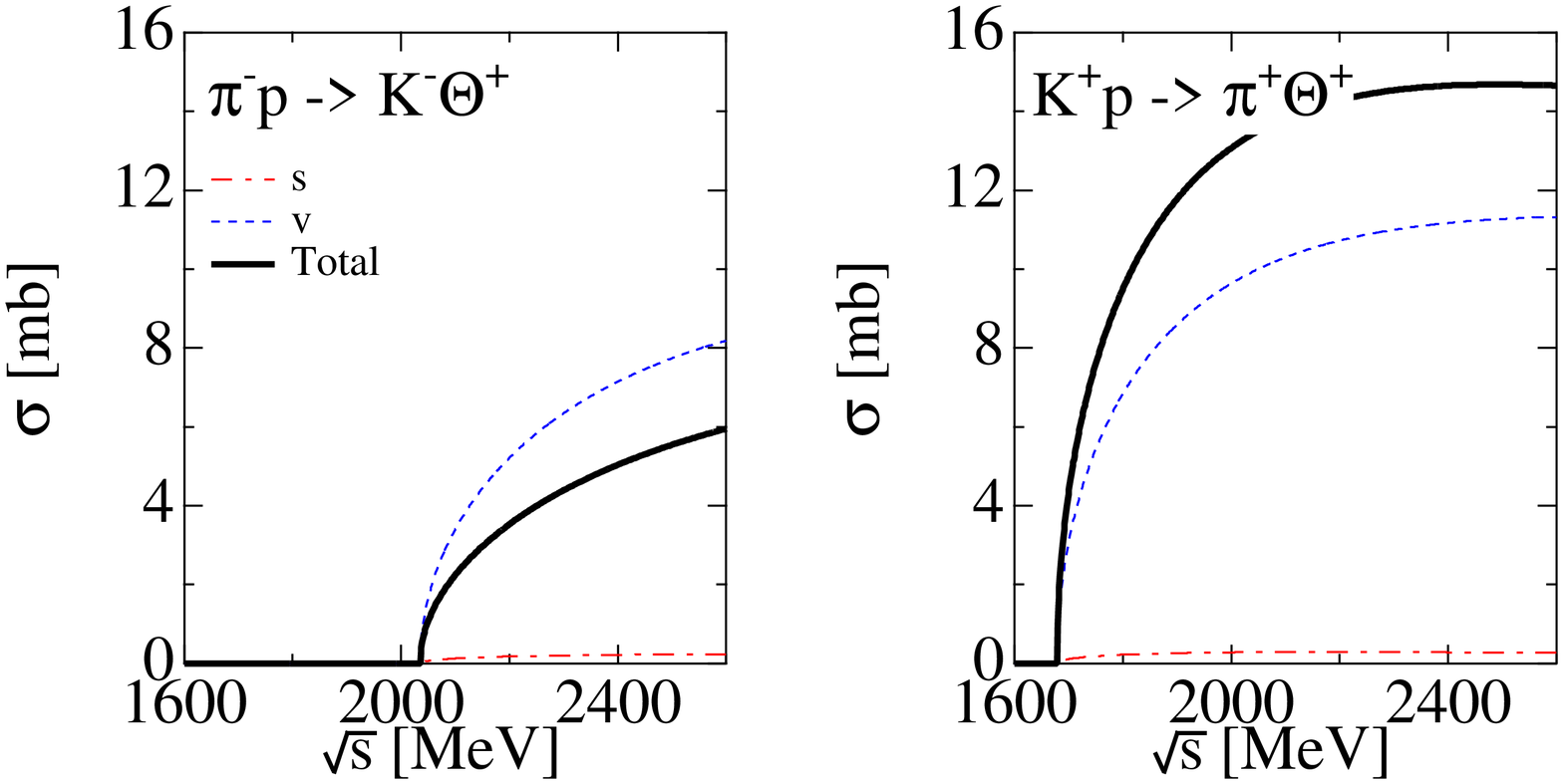}
    \caption{\label{fig:1o2_2}
    (Color online) Total cross sections 
    for the $J^P=1/2^+$ case with $g^s=0.47$ and $g^v=-0.47$.
    The thick line shows the result with full amplitude.
    Dash-dotted and dashed lines are the contributions from
    $s$ and $v$ terms, respectively.}
\end{figure}

As already mentioned, the relative phase of the two coupling constants 
is not determined. If we change the sign,
\begin{equation}
    g^s_{1/2^+} = 0.47,\quad g^v_{1/2^+} = -0.47,
    \label{eq:1o2_2}
\end{equation}
then the results change as in Fig.~\ref{fig:1o2_2}, where constructive
and destructive interference appears in an opposite manner. It is worth
noting that the amplitude for $\piK$ with the relative phase of 
Eq.~\eqref{eq:1o2_1} and that for $\Kpi$ with Eq.~\eqref{eq:1o2_2} are 
the same, as seen in Eq.~\eqref{eq:ampsqure1o2}. The difference only
comes from the kinematic factors in cross section~\eqref{eq:dcross}.

There is a preliminary result from KEK~\cite{ImaiMiwa} that the cross
section of $\piK$ was found to be very small, of the order of a few
$\mu$b. At this stage, we do not want to calculate the cross section
quantitatively, but the experimental result suggests that the relative
phase of Eq.~\eqref{eq:1o2_2} should be plausible, for the small cross
section for the $\piK$ reaction. In this case, the cross section for 
$\Kpi$ becomes large.

\begin{figure}[tbp]
    \centering
    \includegraphics[width=8cm,clip]{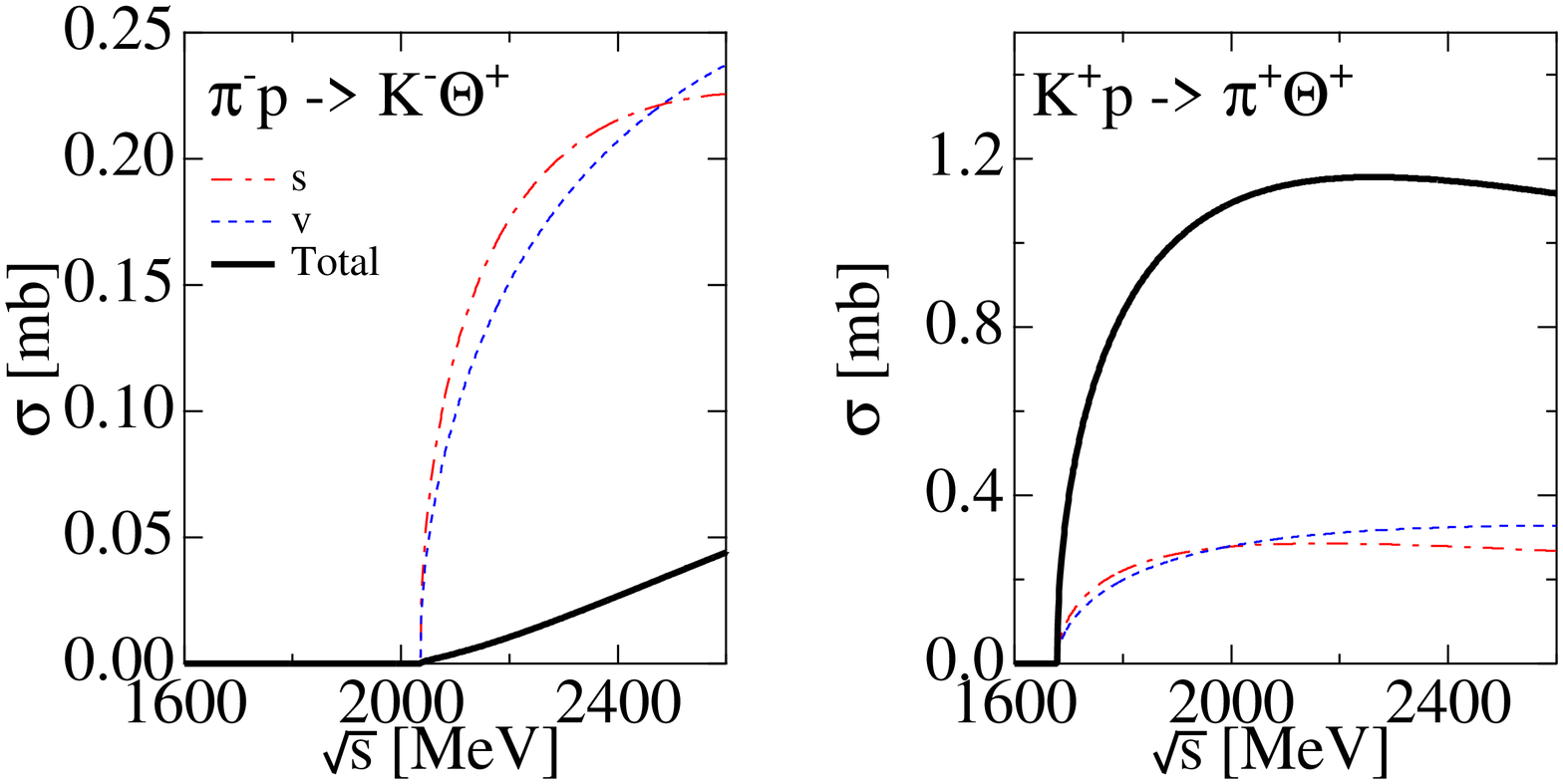}
    \caption{\label{fig:1o2_3}
    (Color online) Total cross sections for the $J^P=1/2^+$ case with
    $g^s=0.47$ and $g^v=-0.08$, when the most destructive interference
    for $\piK$ takes place. Note that the vertical scale is different
    in the two panels. The thick line shows the result with full 
    amplitude. Dash-dotted and dashed lines are the contributions from
    $s$ and $v$ terms, respectively.}
\end{figure}

As a trial, let us search for the set of coupling constants with which
the most destructive interference takes place in $\piK$, by changing
$g^v_{1/2^+}$ within the interval~\eqref{eq:vlimit}. This means that 
the difference between cross sections of $\piK$ and $\Kpi$ is maximal.
Then we find
\begin{equation}
    g^s_{1/2^+} = 0.47,\quad g^v_{1/2^+} = -0.08.
    \label{eq:1o2_3}
\end{equation}
The result is shown in Fig.~\ref{fig:1o2_3}. A huge difference between
$\piK$ and $\Kpi$ can be seen. In this case, we observe the ratio of
cross sections
\begin{equation}
    \frac{\sigma(\Kpi)}{\sigma(\piK)}
    \sim 50 .
    \label{eq:ratio1o2}
\end{equation}
Here we estimated the cross section $\sigma$ as the average of the cross
section shown in the figures (from threshold to 2.6 GeV). Notice that the
ratio of the coupling constants $g^s_{1/2^+}/g^v_{1/2^+} \sim -5.9$ is
relevant for the interference effect. It is possible to scale both 
coupling constants within experimental uncertainties. This does not 
change the ratio of cross sections, but it does change the absolute values.

Next we examine the case with $J^P=3/2^-$. Again, we observe 
constructive and destructive interferences, depending on the relative
sign of the two amplitudes. The interference effect is prominent
around the energy region close to the threshold but is not very strong 
in the higher energy region, compared with $1/2^+$ case.

\begin{figure}[tbp]
    \centering
    \includegraphics[width=8cm,clip]{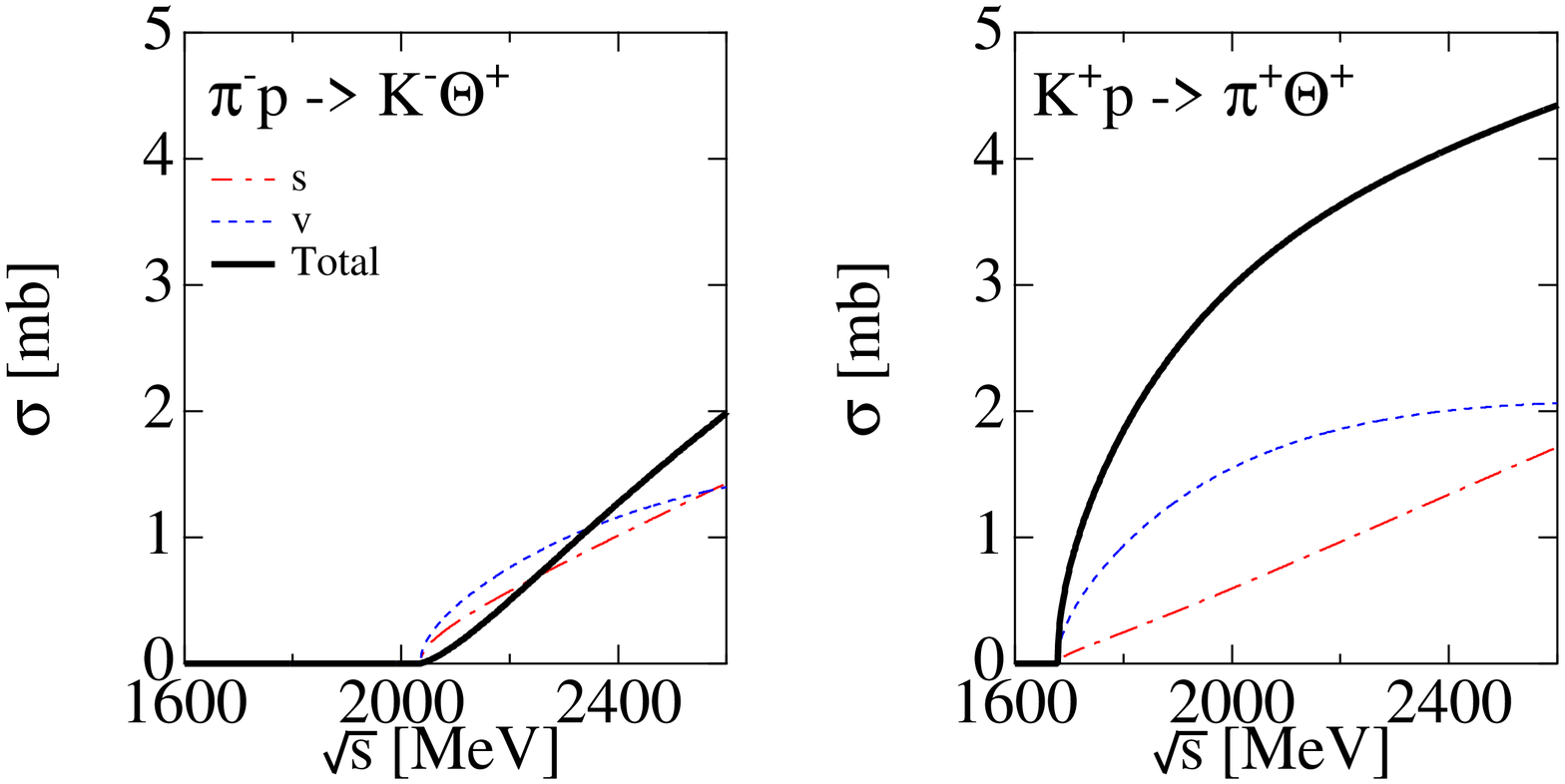}
    \caption{\label{fig:3o2_3}
    (Color online) Total cross section for the $J^P=3/2^-$ case with
    $g^s=0.2$ and $g^v=0.4$, when the most destructive interference for 
    $\piK$ takes place. The thick line shows the result with full amplitude.
    Dash-dotted and dashed lines are the contributions from $s$ and $v$ 
    terms, respectively.}
\end{figure}

We search for the coupling constants with which the most destructive 
interference takes place for $\piK$. We find that destructive 
interference is maximized when the ratio of the coupling constants is
$g^s_{3/2^-}/g^v_{3/2^-} \sim 0.5$. Taking, for instance, the  values
\begin{equation}
    g^s_{3/2^-} = 0.2,\quad g^v_{3/2^-} = 0.4,
    \label{eq:3o2_3}
\end{equation}
which are within the experimental bounds given in Sec.~\ref{sec:coupling},
we obtain the results shown in Fig.~\ref{fig:3o2_3}. In contrast to the
$J^P=1/2^+$ case, here the ratio of cross section is not very large:
\begin{equation}
    \frac{\sigma(\Kpi)}{\sigma(\piK)}
    \sim 3.3 .
    \label{eq:ratio3o2}
\end{equation}
The high-energy behavior in this case is understood from the $p$-wave
nature of the coupling.

Let us mention the effect of the vector meson propagator. For 
simplicity, we take the same value for the coupling constants. First,
we address the $J^P=1/2^+$ case. Without introducing the vector meson
propagator $F(k-k^{\prime})$, the magnitude and energy dependence of 
the vector term is not similar to the scalar one, reflecting the 
structure of amplitudes~\eqref{eq:samp1o2reac} and \eqref{eq:vamp1o2reac}.
The difference between $s$ and $v$ amplitudes is 
$(2\sqrt{s}-M_\Theta-M_p)/2f$ with the same coupling constant, which
ranges from 3 to 14 in the energy region under consideration. The cross
section is proportional to its square, and therefore the vector term
becomes the dominant one. The inclusion of the vector meson propagator
reduces the cross section of the vector term, especially in the 
high-energy region. This eventually leads to the similar energy 
dependence of the two amplitudes $t^s$ and $t^v$, resulting in a large 
cancellation between them, as seen in Fig.~\ref{fig:1o2_3}, although 
a factor $g^s_{1/2^+}/g^v_{1/2^+} \sim -5.9$ is still required to make
the magnitude the same.

For the $J^P=3/2^-$ case, without including the vector meson propagator,
the scalar and vector contributions to the total cross section [the first
and the third terms in Eq.~\eqref{eq:ampsqure3o2}] become exactly the
same, when we take the same coupling constant. Obviously, as seen in
Eq.~\eqref{eq:ampsqure3o2}, the difference of the squared amplitudes is 
the term proportional to $\bm{k}\cdot\bm{k}^{\prime}\propto \cos\theta$,
which goes away when the angular integral is performed. This, however, 
does not lead to complete destructive interference, owing to the second
term in Eq.~\eqref{eq:ampsqure3o2}. The vector meson propagator acts in 
the same way as before, and we obtain somehow a different energy 
dependence of the $s$ and $v$ results (Fig.~\ref{fig:3o2_3}) and a factor 
$g^s_{3/2^-}/g^v_{3/2^-} \sim 0.5$ to compensate for the reduction of 
the cross section of the vector term.

\subsection{Hadronic form factor}

Here we consider the reaction mechanism in detail to give a more 
quantitative result. First we introduce a hadronic form factor at the
vertices, which accounts for the energy dependence of the coupling
constants. Physically, it is understood as the reflection of the finite
size of the hadrons. In practice, however, the introduction of the form
factor has some ambiguities in its form and the cutoff 
parameters~\cite{Nam:2003uf}, which hopefully can be determined from 
experiment.

\begin{figure}[tbp]
    \centering
    \includegraphics[width=8cm,clip]{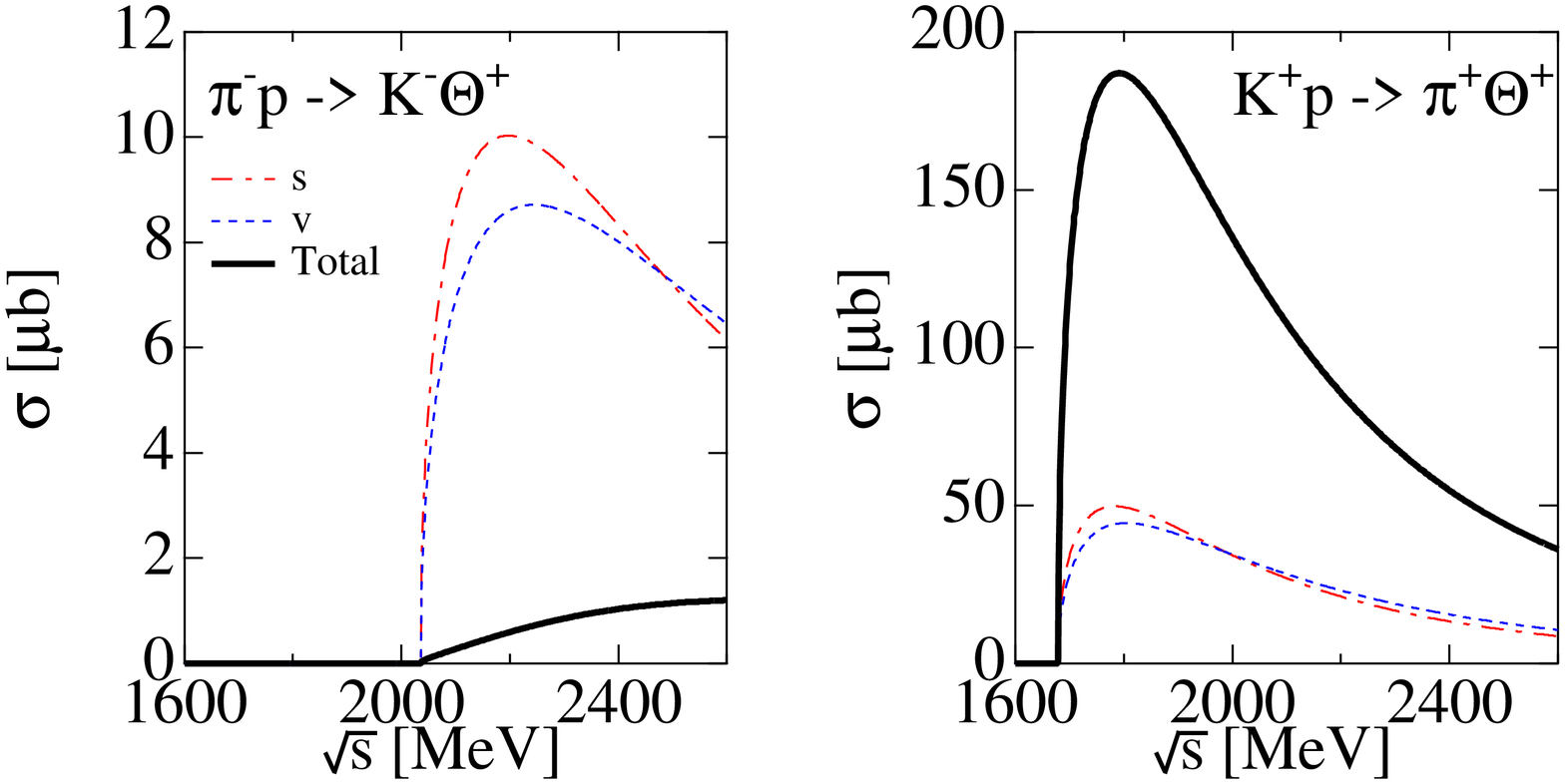}
    \caption{\label{fig:1o2_hFF}
    (Color online) Total cross sections for the $J^P=1/2^+$ case with 
    $g^s=0.47$ and $g^v=-0.08$, including a hadronic form 
    factor~\eqref{eq:monopoleFF}. The thick line shows the result with
    full amplitude. Dash-dotted and dashed lines are the contributions
    from $s$ and $v$ terms, respectively.}
\end{figure}

\begin{figure}[tbp]
    \centering
    \includegraphics[width=8cm,clip]{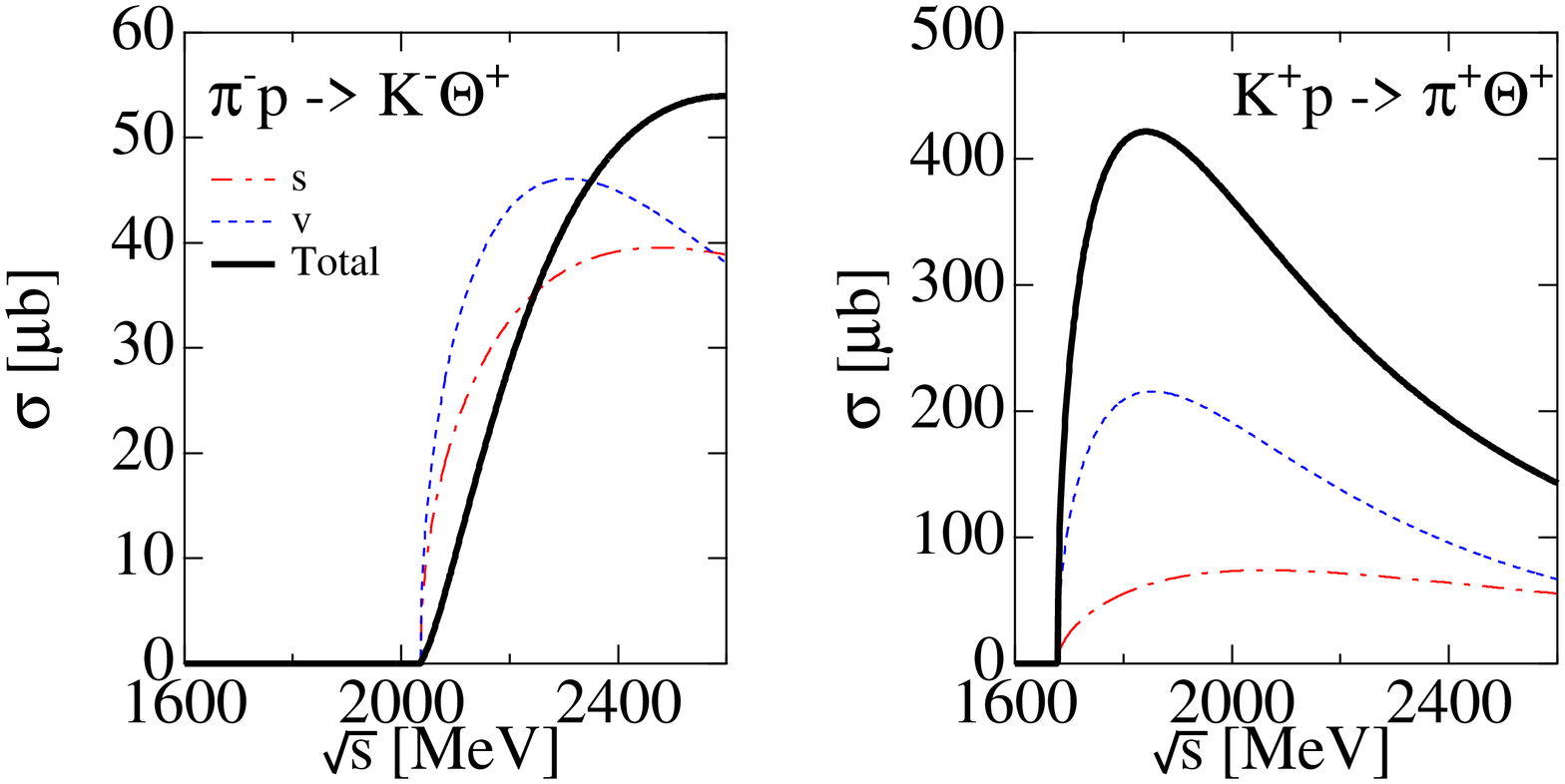}
    \caption{\label{fig:3o2_hFF}
    (Color online) Total cross sections for the $J^P=3/2^-$ case with
    $g^s=0.2$ and $g^v=0.4$, including a hadronic form 
    factor~\eqref{eq:monopoleFF}. The thick line shows the result with 
    full amplitude. Dash-dotted and dashed lines are the contributions 
    from $s$ and $v$ terms, respectively.}
\end{figure}

In Ref.~\cite{Oh:2003kw}, the $\piK$ reaction is studied with a 
three-dimensional monopole-type form factor
\begin{equation}
    F(\sqrt{s})=\frac{\Lambda^2}{\Lambda^2+\bm{q}^2},
    \label{eq:monopoleFF}
\end{equation}
where $\bm{q}^2=\lambda(s,M_N^2,m_{in}^2)/4s$ with $m_{in}$ being the 
mass of the incoming meson and $\Lambda=0.5$ GeV. Here we adopt this form
factor and apply it to the present process. We obtain the results for 
$J^P=1/2^+$ in Fig.~\ref{fig:1o2_hFF} and for $J^P=3/2^-$ in 
Fig.~\ref{fig:3o2_hFF}, with the coupling constants given in 
Eqs.~\eqref{eq:1o2_3} and \eqref{eq:3o2_3}. With this form factor, the
energy of the $\Kpi$ reaction of the ongoing experiment at KEK 
($P_{\text{lab}}\sim 1200$ MeV, $\sqrt{s}\sim 1888$ MeV) is close to
the maximum value for the cross section.

Notice that the ratio of the cross sections of $\piK$ and $\Kpi$ becomes 
larger than those of Eqs.~\eqref{eq:ratio1o2} and \eqref{eq:ratio3o2}.
This is due to the use of the form factor~\eqref{eq:monopoleFF}, which
contains the mass of the initial meson. It further contributes a factor
$\sim 2$ for the ratio of $\piK$ and $\Kpi$.

We observe that the cross section is suppressed down to $\sim 1 \mu$b 
for the $\piK$ reaction in the $1/2^+$ case. However, this is also a 
consequence of our choice of small coupling constants. Indeed, with these
coupling constants, the self-energy of $\Theta^+$ becomes
\begin{equation}
    \re\Sigma_{\Theta^+}^{1/2^+}
    = \re\Sigma^s_{\Theta^+} 
    +\re\Sigma^v_{\Theta^+}
    \sim -5.3-1.6=-6.9 \text{ MeV},
    \label{eq:self1}
\end{equation}
for $p^{0}=1700$ MeV and a cutoff $750$ MeV. This is too small, but as we
mentioned before, we can scale these constants without changing the ratio 
of $\Kpi$ and $\piK$. We would like to search for the coupling constants
which provide a small cross section for $\piK$ reaction compatible with
experiment and a moderate amount of self-energy, which guarantee the
dominance of the two-meson coupling terms compared with the $KN\Theta^+$ 
vertex.

\begin{figure*}[tbp]
    \centering
    \includegraphics[width=8cm,clip]{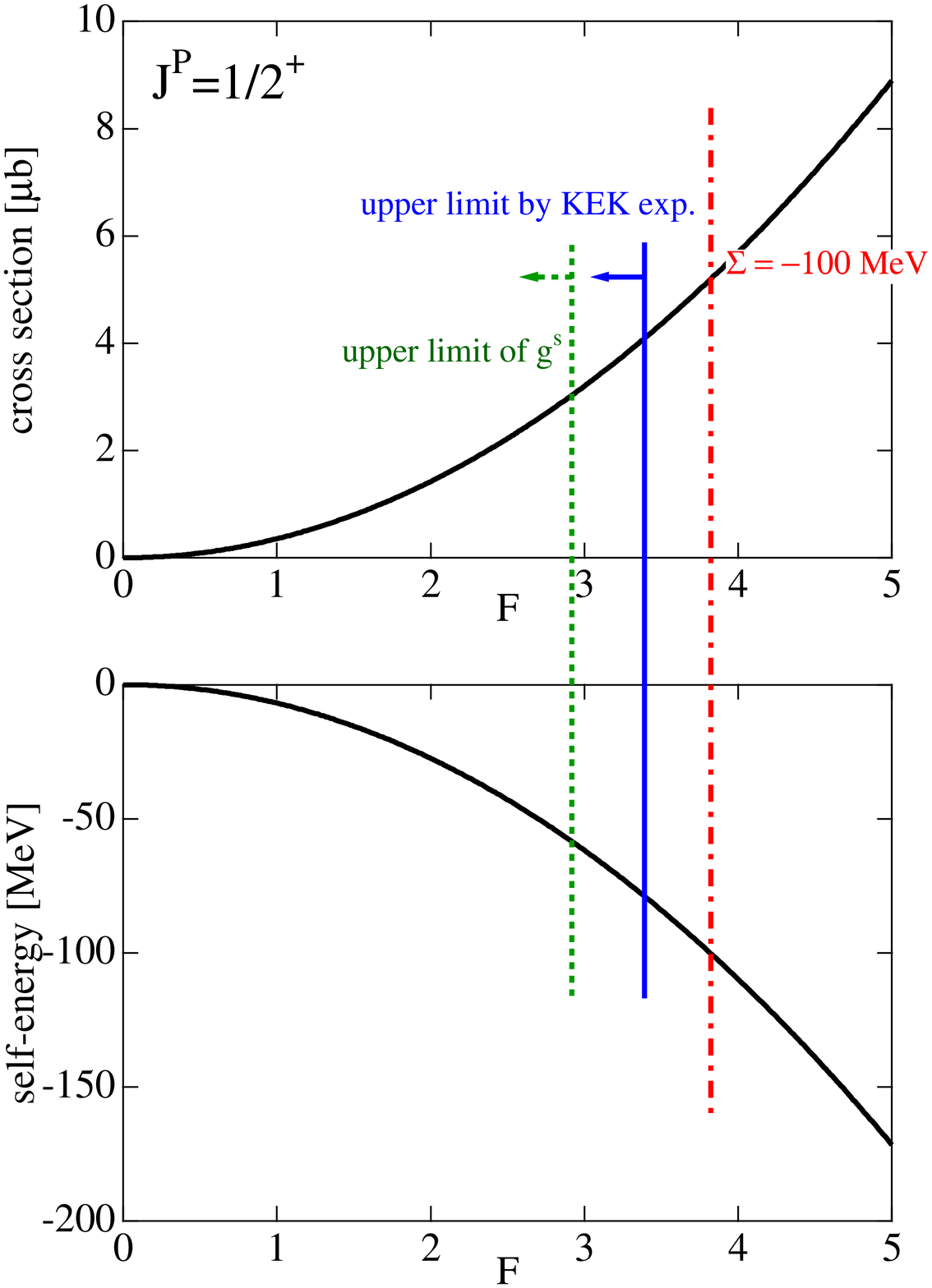}
    \includegraphics[width=8cm,clip]{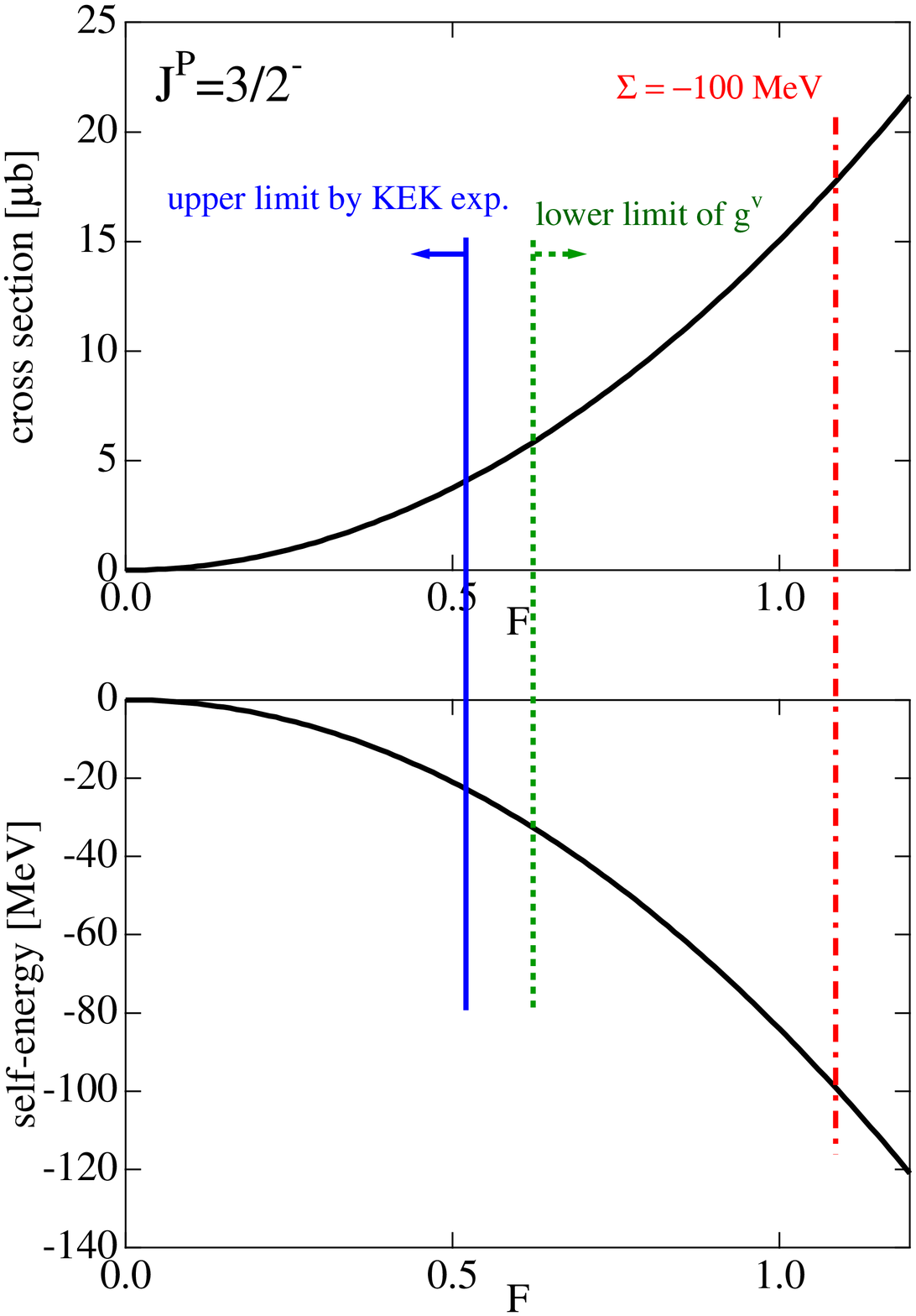}
    \caption{\label{fig:selfcross}
    (Color online) The total cross section of $\piK$ at 
    $P_{\text{lab}}=1920$ MeV and the real part of the self-energy
    of $\Theta^+$ as functions of the factor $F$ defined in 
    Eqs.~\eqref{eq:factor} and \eqref{eq:factor3o2} for $J^P=1/2^+$ (left)
    and $J^P=3/2^-$ (right). Solid, dashed and dash-dotted vertical lines
    show the upper limit of cross section given by the KEK
    experiment~\cite{ImaiMiwa}, the limit of coupling constant, and the
    point where $\re\Sigma=-100$ MeV.}
\end{figure*}

In Fig.~\ref{fig:selfcross}, we plot the cross section of $\piK$ reaction
and the self-energy of $\Theta^+$ by fixing the ratio of coupling 
constants. The cross section is the value at $\sqrt{s}=2124$ MeV, which
corresponds to the KEK experiment $P_{\text{lab}}\sim$ 1920 MeV. The 
horizontal line denotes the factor $F$, which is defined by
\begin{equation}
    g^s_{1/2^+} = F\times 0.47,\quad
    g^v_{1/2^+} = -F\times 0.08 .
    \label{eq:factor}
\end{equation}
We use $F=1$ for the calculation of Fig.~\ref{fig:1o2_hFF}. Both the 
cross section and self-energy are proportional to the square of the
coupling constant. Using the maximum value of cross section 
$\sim 4.1\mu$b~\footnote{The upper limit $\sim 4.1\mu$b is determined 
by assuming the isotropic decay of $\Theta^+\to KN$~\cite{ImaiMiwa}. In
our calculation, the angular dependence of the decay is not very large,
as shown in the following. Therefore we simply use this value for the
estimation of upper limit here.} estimated by KEK 
experiment~\cite{ImaiMiwa}, we have
\begin{align}
    g^s_{1/2^+} &= 1.59, &
    g^v_{1/2^+} &= -0.27,\label{eq:KEK1o2} \\
    \sigma_{\piK}&= 4.1 \mu\text{b}, &
    \re\Sigma_{\Theta} &= -78 \text{ MeV}.
    \nonumber
\end{align}
Furthermore, if we use the upper limit of the scalar term of the coupling
constant, we fix
\begin{align}
    g^s_{1/2^+} &= 1.37, &
    g^v_{1/2^+} &= -0.23, \label{eq:coupling1o2} \\
    \sigma_{\piK} &= 3.2 \mu\text{b},&
    \re\Sigma_{\Theta} &= -58 \text{ MeV}.
   \nonumber
\end{align}
However, if we want to obtain $\re\Sigma_{\Theta}=-100$ MeV, we have
\begin{align}
    g^s_{1/2^+} &= 1.80,&
    g^v_{1/2^+} &= -0.31, \label{eq:1001o2}\\
    \sigma_{\piK} &= 5.0 \mu\text{b},&
    \re\Sigma_{\Theta} &= -100 \text{ MeV}.
    \nonumber
\end{align}
We see that a sizable self-energy is obtained with the coupling constants
\eqref{eq:KEK1o2} and \eqref{eq:1001o2}. These results are summarized in 
Table~\ref{tbl:summary}.

\begin{table}[tbp]
    \centering
    \caption{Summary of the coupling constants, cross sections and 
    self-energies. $\sigma_{\pi^-}$ is the total cross section for $\piK$
    are the values at $P_{\text{lab}}=1920$ MeV; $\sigma_{K^+}$ is that 
    for $\Kpi$, which is the upper limit of the cross section at 
    $P_{\text{lab}}=1200$ MeV.}
    \begin{ruledtabular}
    \begin{tabular}{llllll}
	$J^P$ & $g^s$ & $g^v$ & $\sigma_{\pi^-}$ [$\mu$b]
	& $\sigma_{K^+}$ [$\mu$b]
	& $\re\Sigma_{\Theta}$ [MeV]  \\
	\hline
	$1/2^+$  & 1.59 & $-0.27$
	& \phantom{0}4.1 & $<$1928 & \phantom{0}$-78$  \\
	 & 1.37 & $-0.23$
	 & \phantom{0}3.2 & $<$1415 & \phantom{0}$-58$  \\
	 & 1.80 & $-0.31$ & \phantom{0}5.0 & $<$2506 & $-100$  \\
	$3/2^-$ & 0.104 & 0.209 & \phantom{0}4.1 & $<$\phantom{0}113 
	& \phantom{0}$-23$  \\
	& 0.125 & 0.25 & \phantom{0}5.9 & $<$\phantom{0}162 
	& \phantom{0}$-32$  \\
	 & 0.22 & 0.44 & 18 & $<$\phantom{0}502 & $-100$  \\
    \end{tabular}
    \end{ruledtabular}
    \label{tbl:summary}
\end{table}

For the $J^P=3/2^-$ case, with $g^s_{3/2^-}=0.2$ and $g^v_{3/2^-}=0.4$,
the self-energy of $\Theta^+$ becomes
\begin{equation}
    \re\Sigma_{\Theta^+}^{3/2^-}
    = \re\Sigma^s_{\Theta^+}
    +\re\Sigma^v_{\Theta^+} 
    \sim -4-80=-84 \text{ MeV}.
    \label{eq:self2}
\end{equation}
In Fig.~\ref{fig:selfcross}, we plot the cross section of the $\piK$ 
reaction and the self-energy of $\Theta^+$ by fixing the ratio of
coupling constants. The horizontal line denotes the factor $F$, which is
defined by
\begin{equation}
    g^s_{1/2^+} = F\times 0.2,\quad
    g^v_{1/2^+} = F\times 0.4.
    \label{eq:factor3o2}
\end{equation}
Using the maximum value of cross section $\sim 4.1\mu$b estimated by 
the KEK experiment, we have
\begin{align}
    g^s_{3/2^-} &= 0.104, &
    g^v_{3/2^-} &= 0.209, \label{eq:KEK3o2} \\
    \sigma_{\piK} &= 4.1 \mu\text{b},&
    \re\Sigma_{\Theta} &= -23 \text{ MeV}.
    \nonumber
\end{align}
In the region plotted in the figure, the coupling constants do not exceed
the upper bounds, but the lower limit of $g^v_{3/2^-}$ appears:
\begin{align}
    g^s_{3/2^-} &= 0.125, &
    g^v_{3/2^-} &= 0.25,\label{eq:coupling3o2}\\
    \sigma_{\piK} &= 5.9 \mu\text{b},&
    \re\Sigma_{\Theta} &= -33 \text{ MeV}.
    \nonumber 
\end{align}
To make $\re\Sigma_{\Theta}=-100$ MeV, we have
\begin{align}
    g^s_{3/2^-} &= 0.22,&
    g^v_{3/2^-} &= 0.44, \label{eq:1003o2} \\
    \sigma_{\piK} &= 18 \mu\text{b},&
    \re\Sigma_{\Theta} &= -100 \text{ MeV}.
    \nonumber
\end{align}

\begin{figure}[tbp]
    \centering
    \includegraphics[width=8cm,clip]{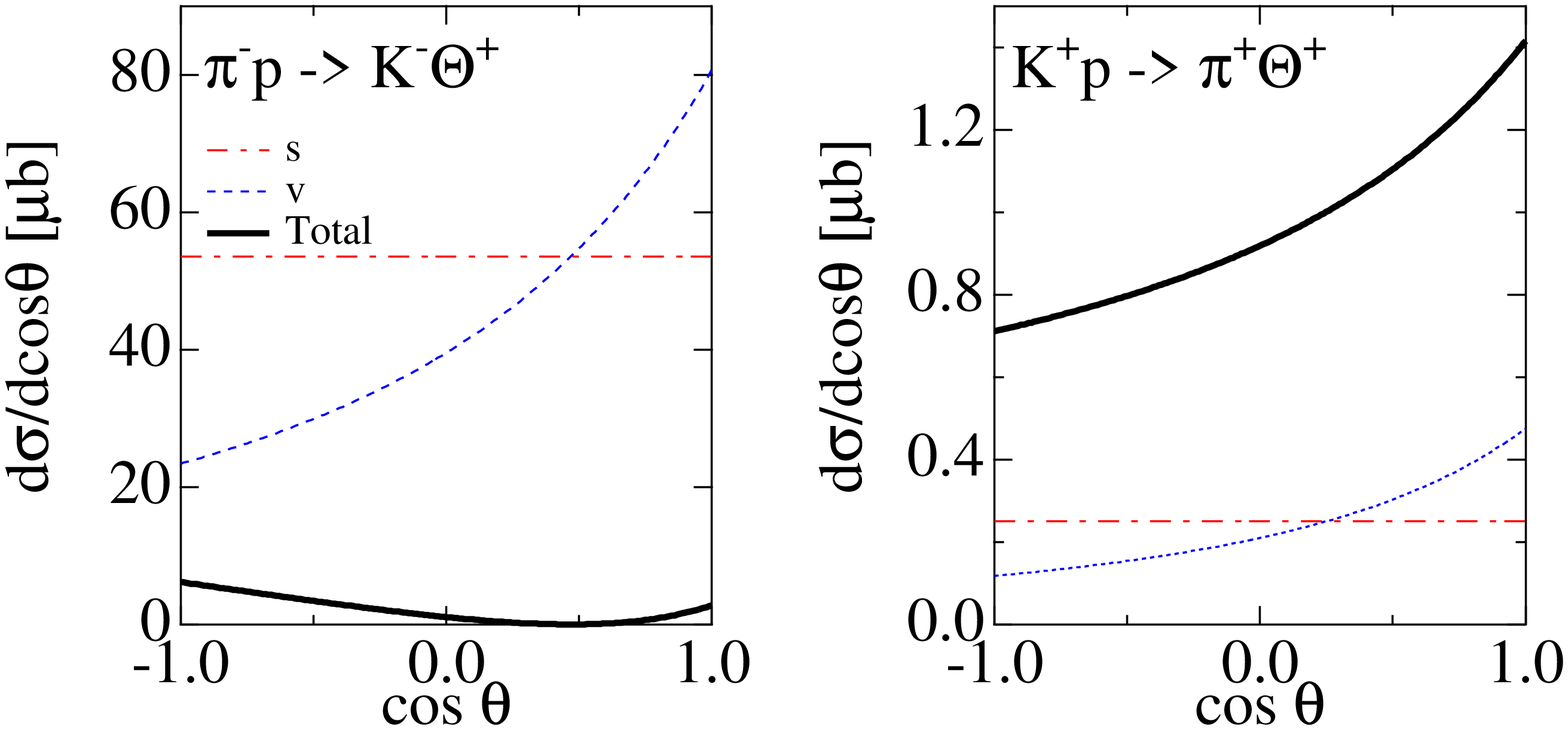}
    \caption{\label{fig:anglecross1o2}
    (Color online) Angular dependence of the differential cross section
    for $\piK$  at $P_{\text{lab}}\sim 1920$ MeV (left) and for $\Kpi$ 
    at $P_{\text{lab}}\sim 1200$ MeV (right) for the $J^P=1/2^+$ case
    with $g^s=1.59$ and $g^v=-0.27$, including a hadronic form 
    factor~\eqref{eq:monopoleFF}. The thick line shows the result with 
    full amplitude. Dash-dotted and dashed lines are the contributions
    from $s$ and $v$ terms, respectively.}
\end{figure}

\begin{figure}[tbp]
    \centering
    \includegraphics[width=8cm,clip]{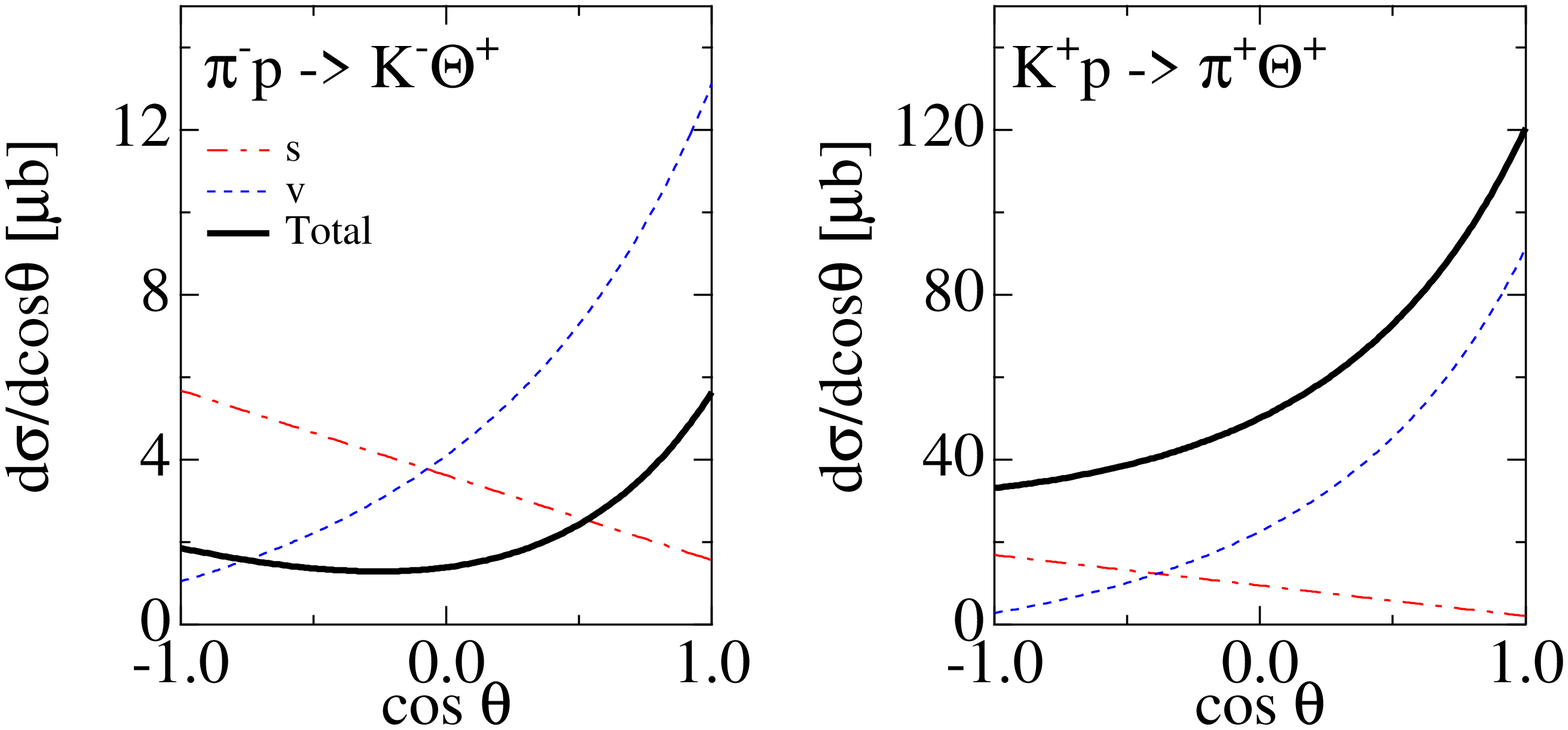}
    \caption{\label{fig:anglecross3o2}
    (Color online) Angular dependence of the differential cross section
    for $\piK$  at $P_{\text{lab}}\sim 1920$ MeV (left) and for $\Kpi$ 
    at $P_{\text{lab}}\sim 1200$ MeV (right) for the $J^P=3/2^-$ case
    with $g^s=0.106$ and $g^v=0.212$, including a hadronic form 
    factor~\eqref{eq:monopoleFF}. The thick line shows the result with
    full amplitude. Dash-dotted and dashed lines are the contributions
    from $s$ and $v$ terms, respectively.}
\end{figure}

Finally, we show the angular dependence of the cross sections. In 
Figs.~\ref{fig:anglecross1o2} and \ref{fig:anglecross3o2}, we plot the 
angler dependence of the differential cross sections at the energy of 
KEK experiment: $P_{\text{lab}}\sim 1920$ MeV for $\piK$ and 
$P_{\text{lab}}\sim 1200$ MeV for $\Kpi$. For the $J^P=1/2^+$ case, the
contribution from the $s$ term has no angular dependence, whereas the $v$
term shows a forward peak, owing to the $t$-channel exchange of the vector
meson propagator. Because of the interference of the two amplitudes, the 
total result becomes zero at $\cos\theta\sim 0.5$ for the $\piK$ reaction.
For the $J^P=3/2^-$ case, the $s$ term varies linearly in $\cos\theta$, 
leading to a backward peak. The $v$ term shows a forward peak, which is 
enhanced by the vector meson propagator. Interference of the two amplitude
leads to a clear forward peak for both $\piK$ and $\Kpi$ reactions.

\subsection{Effect of Born terms}

\begin{figure}[tbp]
    \centering
    \includegraphics[width=4cm,clip]{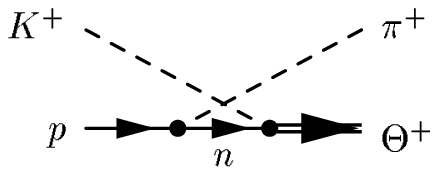}
    \includegraphics[width=4cm,clip]{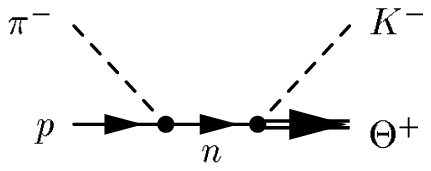}
    \caption{\label{fig:Born}
    Born terms for the reaction: $u$-channel diagram for 
    $\Kpi$ (left); $s$-channel diagram for
    $\piK$ (right).}
\end{figure}

In this subsection, we briefly discuss the possible effect from the Born
terms, as shown in Fig.~\ref{fig:Born}, which have not been taken into
account in the present studies. However, there are reasons that the Born
terms are not important in the present reactions. First, the Born terms
are proportional to the decay width of $\Theta^+$ and therefore suppressed
if the decay width of the $\Theta^+$ is narrow. Second, in the energy 
region of $\Theta^+$ production, the energy denominator of the exchanged 
nucleon suppresses the contribution, especially for the $s$-channel term
in the $\piK$ reaction. Here we would like to confirm this explicitly.

At the tree level, there are $s$-, $t$-, and $u$-channel diagrams. 
However, assuming $I=0$ for $\Theta^+$, there is only a $u$ channel in 
$\Kpi$ (Fig.~\ref{fig:Born}, left), whereas there is only an $s$ channel 
in $\piK$ (Fig.~\ref{fig:Born}, right). For these terms, we need the
Yukawa couplings such as $KN\Theta^+$ and $\pi NN$ couplings. There are
two schemes to introduce the Yukawa couplings, namely, pseudoscalar (PS) 
and pseudovector (PV) schemes. For the construction of the Born amplitude,
it is reasonable to rely on chiral symmetry, where the two schemes should
provide the same results.  

In this case the meson-baryon scattering amplitude should be a quantity
of ${\cal O}(k)$ or higher, where $k$ is a momentum of mesons. In the PV 
scheme, since each $KN\Theta$ coupling is of ${\cal O}(k)$, the Born 
amplitude behaves as ${\cal O}(k^2)$, consistent with this observation.  
In contrast, a naive construction of the Born term in the PS scheme 
leads to an amplitude of ${\cal O}(1)$. It is well-known that a scalar
exchange term cancel the term of ${\cal O}(1)$. However, the interaction 
of the scalar channel is not well understood. Therefore, we adopt the PV
scheme in the following study. Another advantage of the PV scheme is that
it can be extended easily to the $J^P=3/2^-$ case, while it is not so 
simple in the PS scheme~\cite{Nam:2005uq}. In this respect, our method 
differs from the previous study of similar reactions~~\cite{Oh:2003kw,
Oh:2003gj}, in which the PS scheme was used.

\begin{figure}[tbp]
    \centering
    \includegraphics[width=8cm,clip]{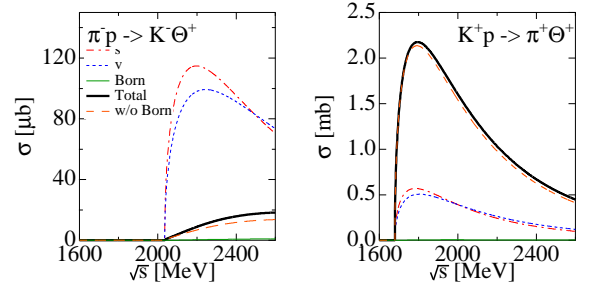}
    \caption{\label{fig:1o2_B2}
    (Color online) Total cross section for the $J^P=1/2^+$ case with 
    $g^s=1.59$ and $g^v=-0.27$, including hadronic form 
    factor~\eqref{eq:monopoleFF} and Born terms. The thick line shows 
    the result with full amplitude. Dash-dotted, dashed, and thin lines 
    are the contributions from $s$, $v$, and Born terms, respectively.
    The result without Born terms are shown by the long-dashed
    lines.}
\end{figure}

The interaction Lagrangians for $1/2^+$ are 
\begin{align}
    \mathcal{L}_{KN\Theta}^{1/2^+}
    &=\frac{g_{A}^{*,1/2^+}}{2f}\bar{\Theta}^+\gamma_{\mu}\gamma_5
    \partial^{\mu}KN + \text{h.c.},
    \label{eq:PVcouplingKNTheta} \\
    \mathcal{L}_{\pi NN}
    &=\frac{g_{A}}{2f}\bar{N}\gamma_{\mu}\gamma_5 
    \partial^{\mu}\bm{\pi} N.
    \label{eq:PVcouplingpiNN}
\end{align}
The fields $N$ and $\bm{\pi}$ are defined in Eq.~\eqref{eq:fielddef},
and the Kaon field is defined as
\begin{equation}
    K=
    \begin{pmatrix}
	-K^0 &
	K^+
    \end{pmatrix},
    \label{eq:Kdef}
\end{equation}
and the coupling constants are determined as
\begin{equation}
    g_{A}^{*,1/2^+} = 0.0935,
    \label{eq:couplingKNTheta}
\end{equation}
which is determined by $\Gamma_{\Theta^+}=1$ MeV, and we use
\begin{equation}
    g_{A} = 1.25.
    \label{eq:couplingpiNN}
\end{equation}
The amplitude for $\pi^-(k)p(p)\to K^-(k^{\prime})\Theta^+(p^{\prime})$
is given by
\begin{align}
    -it
    &= i\sqrt{2}\frac{g_{A}^{*,1/2^+}g_A}{4f^2}
    (\bm{\sigma}\cdot\bm{k}^{\prime})
    \frac{M}{E}\frac{1}{p_0+k_0-E(\bm{p}+\bm{k})}
    (\bm{\sigma}\cdot\bm{k})
    \nonumber ,
\end{align}
and for $K^+(k)p(p)\to \pi^+(k^{\prime})\Theta^+(p^{\prime})$,
\begin{align}
    -it
    &= i\sqrt{2}\frac{g_{A}^{*,1/2^+}g_A}{4f^2}
    (\bm{\sigma}\cdot\bm{k})
    \frac{M}{E}\frac{1}{p_0-k_0^{\prime}-E(\bm{p}-\bm{k}^{\prime})}
    (\bm{\sigma}\cdot\bm{k}^{\prime}) 
    \nonumber .
\end{align}
In Fig.~\ref{fig:1o2_B2} we show the results including Born terms.
We can observe that the effect of Born terms is indeed small
in both reactions.

For the $J^P=3/2^-$ case, the interaction Lagrangian can be written as
\begin{align}
    \mathcal{L}_{KN\Theta}^{3/2^-}
    &=\frac{g_{A}^{*,3/2^-}}{2f}\bar{\Theta}^{+
    \mu}\gamma_5
    \partial_{\mu}KN + \text{h.c.},
    \label{eq:PVcouplingKNTheta3o2}
\end{align}
with the same $\pi NN$ vertex in Eq.~\eqref{eq:couplingpiNN}. In the 
nonrelativistic expansion this term yields a $d$-wave coupling  so that
the square of momenta appears in the vertex. It reduces the contribution 
further than the $1/2^+$ case, and therefore, the effect of Born terms 
for $J^P=3/2^-$ is also small.

\section{Summary}\label{sec:summary}

In this paper, we studied the two-meson couplings of $\Theta^+$ for
$J^P=1/2^+$ and $3/2^-$. The effective interaction Lagrangians for the
two-meson coupling were given, and these coupling constants were 
determined based on the $\bm{8}$-$\overline{\bm{10}}$ representation 
mixing scheme, by using information of the $N^*\to \pi\pi N$ decays.
These values were further constrained in order to provide appropriate 
size of the self-energy of the $\Theta^+$. Finally, we applied the 
effective Lagrangian to the meson induced reactions $\piK$ and $\Kpi$.

We found that there is an interference effect between the two amplitudes
of the scalar and vector types, which can help to explain the very small 
cross section for the $\piK$ reaction observed at KEK~\cite{ImaiMiwa}.
In this case, reflecting the symmetry under exchange of two amplitudes,
the large cross section for $\Kpi$ reaction was obtained as a consequence
of the interference. The interference occurs in both $1/2^+$ and $3/2^-$
cases.

In Table~\ref{tbl:summary}, we summarize the results obtained in the
present analysis. For a given set of coupling constants, the upper bound
of the cross section of the $\Kpi$ reaction is estimated by maximizing
the interference effect. We observe that the large cross section of the
order of millibarns for $\Kpi$ is obtained for the $1/2^+$ case, whereas 
the upper limit of the cross section is not very large for $3/2^-$ case.
Therefore, if large cross sections are observed in the $\Kpi$ reaction,
it would indicate $J^P=1/2^+$ for the $\Theta^+$.

For completeness, we would like to mention the case where the cross
sections for both $\piK$ and $\Kpi$ reactions are small. If the cross 
section of $\Kpi$ reaction is also small, it could be explained by the
small coupling constants and is not an interference effect. For the 
$J^P=1/2^+$ case, both coupling constants can be zero within the
experimental uncertainties. However, for the $3/2^-$ case, there is a
lower limit for the $g_{3/2^-}^v$, which means that the lower limit is
also imposed for the cross sections. We search for the set of coupling 
constants that provide the minimum value for the $\Kpi$ cross section,
keeping a $\piK$ cross section to be less than $4.1 \mu$b. We obtain 
$\sigma_{\Kpi}\sim 58 \mu$b with $g_{3/2^-}^s=0.04$ and $g_{3/2^-}^v=0.18$.
However, one should notice that the small coupling constants do not 
guarantee the dominance of two-meson coupling, and the Born terms and 
interference effect may play a role, which is beyond our present scope.

The present analysis provides an extension of effective interactions
obtained in Ref.~\cite{Cabrera:2004yg} with representation mixing and
$J^P=3/2^-$. It is also interesting to apply the present extension to the
study of the medium effect of $\Theta^+$~\cite{Cabrera:2004yg} and the
production of $\Theta^+$ hypernuclei~\cite{Nagahiro:2004wu}. From the
experimental point of view, the cross section of $\Kpi$ reaction is of
particular importance to the present results. To perform a better analysis
for the two-meson coupling, more experimental data for three-body decays
of nucleon resonances are strongly desired.
 
\begin{acknowledgments}
    We acknowledge Manuel J. Vicente Vacas, Eulogio Oset, Seung-il Nam, 
    and Koji Miwa for useful discussion and comments. One of the authors
    (T.~H.) thanks the Japan Society for the Promotion of Science (JSPS)
    for support. This work is supported in part by the Grant for 
    Scientific Research [(C) No.17959600, T.~H.] and [(C) No.16540252,
    A.~H.] from the Ministry of Education, Culture, Science and 
    Technology, Japan.
\end{acknowledgments}

\end{document}